\documentclass[a4paper, 10pt]{article}
\usepackage{jheppub} 
\usepackage{hyperref}
\usepackage[utf8x]{inputenc} 
\usepackage[dvipsnames]{xcolor}
\usepackage{booktabs}
\usepackage{arydshln}

\DeclareMathAlphabet{\mathsfit}{\encodingdefault}{\sfdefault}{m}{sl}

\def\met{E_T^\mathrm{miss}}
\def\ptmiss{\vec{p}_T^{\text{miss}}}

\definecolor{darkred}{rgb}{0.6, 0, 0}
\definecolor{darkblue}{rgb}{0, 0, 0.6}
\hypersetup{colorlinks=true, linkcolor=darkblue, urlcolor=darkblue, citecolor=darkred}

\title{Deep learning approaches to top FCNC couplings to photons at the LHC}

\author[a]{Benjamin Fuks,}
\author[b]{\! Sumit K.~Garg,}
\author[c]{\! A.~Hammad}
\author[d,e]{\! and Adil Jueid}

\affiliation[a]{ Laboratoire de Physique Th\'{e}orique et Hautes \'{E}nergies (LPTHE), UMR 7589,\\ Sorbonne Universit\'{e} \& CNRS, 4 place Jussieu, 75252 Paris Cedex 05, France}\emailAdd{fuks@lpthe.jussieu.fr}
\affiliation[b]{Manipal Centre for Natural Sciences, Manipal Academy of Higher Education, Dr.T.M.A. Pai Planetarium Building, Manipal-576104, Karnataka, India}\emailAdd{sumit.kumar@manipal.edu}
\affiliation[c]{Theory Center, IPNS, KEK,  1-1 Oho, Tsukuba, Ibaraki 305-0801, Japan}\emailAdd{hamed@post.kek.jp}
\affiliation[d]{Particle Theory and Cosmology Group, Center for Theoretical Physics of the Universe, \\ Institute for Basic Science (IBS), Daejeon, 34126, Republic of Korea} \emailAdd{adiljueid@ibs.re.kr}
\affiliation[e]{Cosmology, Gravitation and Astroparticle Physics Group, Center for Theoretical Physics of the Universe, \\ Institute for Basic Science (IBS), Daejeon, 34126, Republic of Korea}
\begin{document}

\abstract{We investigate the sensitivity of the LHC to flavour-changing neutral current interactions involving the top quark and a photon using a model-independent effective field theory framework, focusing on two complementary processes: single top production via $qg \to t\gamma$ and the rare decay $t \to q\gamma$ in top pair events. To enhance signal discrimination, we employ a range of deep learning classifiers, including multi-layer perceptrons, graph attention networks and transformers, and compare them against a traditional cut-based analysis. Our results demonstrate that attention-based architectures, in particular transformer networks, significantly outperform other strategies, yielding up to a factor of five improvement in the expected exclusion limits. In particular, we show that at the high-luminosity LHC, rare top branching ratios can be probed down to values as low as $10^{-6}$. Our results thus highlight the significant potential of attention-based architectures for improving the sensitivity to new physics signatures in top quark processes at colliders. }

\maketitle
%\tableofcontents

%%%%%%%%%%%%%%%
\section{Introduction}
\label{S:Intro}
%%%%%%%%%%%%%%%
The discovery of CP violation in kaon systems played a crucial role in motivating the existence of the top quark, highlighting the requirement for a third family of quarks to enable a CP-violating phase in the Cabibbo-Kobayashi-Maskawa (CKM) matrix. However, the experimental confirmation of the top quark was achieved only in 1995 by the CDF and D0 experiments at Fermilab~\cite{CDF:1995wbb, D0:1995jca}. Thirty years later, the top quark is recognised as the heaviest among all known elementary particles with a mass that originates, in the Standard Model (SM), from the top interaction with the Higgs field and the related sizeable Yukawa coupling. This suggests that the top quark has a special role in electroweak symmetry breaking, and that it is possibly connected to yet-unknown new dynamics. It is therefore widely regarded as a powerful probe for physics beyond the SM (BSM). Consequently, measurements involving top quark production and decay processes, as well as precisely measured observables where the top contributes via quantum loops, are deeply scrutinised as they can potentially provide vital information about BSM scenarios~\cite{Atwood:2000tu}.

With the ongoing third run of the LHC, both the CMS and ATLAS experiments will soon have access to about 500~fb$^{-1}$ of data. This large integrated luminosity opens the possibility to search for rare phenomena corresponding to production cross sections of only a few fb. Among all these rare processes of potential interest, those depending on flavour-changing neutral current (FCNC) interactions involving the top quark are particularly relevant. In the SM, top FCNCs are highly suppressed due to the Glashow-Iliopoulos-Maiani (GIM) mechanism, leading to extremely small branching fractions for decays such as $t \to qZ$, $t \to q H$, $t \to q\gamma$ and $t \to qg$ with $q = u, c$, typically in the range $10^{-17}-10^{-12}$~\cite{Eilam:1990zc, Mele:1998ag, Aguilar-Saavedra:2002lwv, Aguilar-Saavedra:2004mfd, Zhang:2013xya, Durieux:2014xla}. While these rates lie far beyond the reach of current experiments, various extensions of the SM can enhance them by several orders of magnitude, sometimes up to $\mathcal{O}(10^{-7})-\mathcal{O}(10^{-4})$~\cite{Castro:2022qkg, Durieux:2014xla, Larios:2006pb, Barros:2019wxe}, bringing them within the expected sensitivity of the LHC experiments. Any observation of a deviation compatible with FCNC effects in the top sector would therefore constitute a strong indication of BSM physics associated with flavour, as typically expected, for instance, in two-Higgs-doublet models~\cite{Atwood:1996vj, Botella:2015hoa, Abbas:2015cua, Baum:2008qm, Balaji:2020qjg}, the minimal supersymmetric standard model~\cite{Dedes:2014asa, Cao:2007dk, Lopez:1997xv, Eilam:2001dh}, left-right (super)symmetric~\cite{Gaitan:2004by, Frank:2005vd, Frank:2023fkc} models,  extra dimensional models~\cite{Gao:2013fxa, Dey:2016cve, Diaz-Furlong:2016ril}, dark matter~\cite{Agrawal:2014aoa, Agrawal:2015kje, Jueid:2024cge} and composite models~\cite{Balaji:2021lpr, Crivellin:2022fdf}.

Top quark FCNCs are typically studied at colliders via two main analysis strategies: top pair production followed by a rare FCNC decay of one of the two produced top quarks, and FCNC single top production followed by a standard top decay. To date, no significant deviation from the SM expectations has been observed in searches performed at LEP~\cite{L3:2002hbp, OPAL:2001spi, ALEPH:2002wad}, HERA~\cite{ZEUS:2003vfj, ZEUS:2011mya} and the Tevatron~\cite{Kikuchi:2000sv, D0:2010dry, D0:2007wfn}. At the LHC, both the ATLAS and CMS collaborations have carried out dedicated searches in multiple channels, and current bounds on branching ratios such as $\text{BR}(t \to qZ)$, $\text{BR}(t \to q\gamma)$, $\text{BR}(t \to qH)$ and $\text{BR}(t \to qg)$ are now typically in the range $10^{-5} - 10^{-3}$~\cite{ATLAS:2023ujo, ATLAS:2023qzr, ATLAS:2022gzn, ATLAS:2021amo, ATLAS:2018zsq, ATLAS:2018jqi, ATLAS:2018xxe, CMS:2016uzc, CMS:2017wcz, CMS:2021hug, CMS:2020utv, 
ATLAS:2024mih, CMS:2024ubt, ATLAS:2015vhj, CMS:2013knb, CMS:2015kek,ATLAS:2022per,CMS:2023bjm, ATLAS:2015iqc,ATLAS:2012lhi}. While they lie still several orders of magnitude above SM expectations, these results can already be used to constrain a variety of new physics models. Furthermore, with the high-luminosity phase of the LHC (HL-LHC) at the horizon, sensitivity to these rare decay processes is expected to significantly improve due to increased luminosity and better detector performance, thus potentially probing (or even ruling out) broad classes of BSM scenarios. 
In light of the anticipated HL-LHC results on top FCNCs, it hence becomes timely to revisit the analysis strategies used to search for these rare processes. 

In this work, we investigate the potential of modern collider analysis methodologies to boost sensitivity to FCNC signals in top quark events. Our goal is to compare several deep learning strategies that we apply to a set of low-level and high-level reconstructed kinematic variables. Deep learning methods often outperform traditional cut-based approaches due to their ability to learn complex and non-linear decision boundaries in high-dimensional feature spaces. % While cut-based analyses rely on manually-defined thresholds applied to a limited set of kinematic variables, thus potentially overlooking important correlations, deep learning techniques can automatically extract and integrate information from all input features.
This subsequently enables them to identify subtle patterns that can be used to distinguish signal from background, resulting in significantly improved classification performance. % In particular, 
We consider three distinct deep learning architectures that have shown promising performance in recent studies: multi-layer perceptrons (MLPs)~\cite{Rumelhart:1986, Goodfellow:2016}, graph attention networks (GATs)~\cite{Velickovic:2017lzs, Yu:2023juh, Calafiura:2024qhv, Esmail:2023axd} and transformer-based models~\cite{Qu:2022mxj, Wu:2024thh, Hammad:2023sbd, Hammad:2024cae}. These architectures differ in how they handle the input data structure and exploit correlations among observables, offering a broad perspective on the role of representation learning in collider physics. GATs and transformer architectures offer significant advantages over MLPs in classification tasks, especially when handling structured or relational data. GATs are particularly well-suited for capturing interactions between reconstructed final-state particles within graph-based event representations, effectively modelling complex topologies and local connectivity patterns that MLPs are not inherently designed to process. Transformers, by contrast, employ self-attention mechanisms to dynamically evaluate the importance of each input particle, assigning context-dependent weights that help distinguish between signal and background. Unlike MLPs, which treat input features as fixed-size unstructured vectors, both GATs and transformers can adapt to the underlying event structure, yielding more expressive models and enhanced classification performance. By comparing their performance across realistic simulated FCNC signal and background samples, we aim to highlight both the strengths and limitations of these approaches and identify possible paths forward for future experimental searches at the LHC.

To provide quantitative estimates of the potential improvements in sensitivity, we work within the framework of the SM effective field theory (SMEFT) which offers a model-independent parametrisation of new physics effects via higher-dimensional operators that respect the SM symmetries~\cite{Buchmuller:1985jz, Grzadkowski:2010es, Brivio:2017vri}. This bottom-up approach is particularly well motivated when possible new particles lie at scales beyond the current experimental reach, and then allows for a consistent and unified treatment of top-quark FCNC processes in both production and decay channels. For practical purposes and in order to streamline the analysis, we describe the relevant BSM interactions in terms of a minimal set of effective operators defined in the broken electroweak phase~\cite{Aguilar-Saavedra:2008nuh}. Furthermore, as a case study we concentrate on processes involving an anomalous coupling between the top quark, a photon and a light up-type quark ($u$ or $c$), and we simulate both the related single-top and top-antitop production channels at the LHC. The performance of the different explored analysis strategies is next evaluated and compared in terms of sensitivity to the FCNC signal considered for a luminosity of 500~fb$^{-1}$. Moreover, we also confront our predictions with existing LHC Run 2 data, normalising our results to an integrated luminosity of 139~fb$^{-1}$ to better assess quantitatively the potential impact of the proposed methods. Finally, although our analysis is applicable to other top FCNC couplings, different collider luminosities and centre-of-mass energies, the associated comprehensive exploration is deferred to future work.

The rest of this report is organised as follows. In section~\ref{S:model}, we present our effective parametrisation of top-quark FCNC interactions and summarise the current experimental constraints from a broad range of experiments. In section~\ref{S:topFCNC}, we introduce our simulation framework and provide details on the generation of both background and signal events in the case of a top FCNC interaction with the photon. Section~\ref{S:methods} outlines the various analysis strategies employed in this study. %, including both a conventional cut-based approach and three different machine learning methods. 
In section~\ref{S:results:tqa}, we report our results for top FCNC interactions involving a photon and a light up-type quark at LHC luminosities of 139 and 500~fb$^{-1}$. We finally summarise our findings and conclude in section~\ref{S:conclusions}.

%%%%%%%%%%%%%%%
\section{Top-quark FCNC interactions at colliders: Lagrangians and bounds}
\label{S:model}
%%%%%%%%%%%%%%%

At high-energy colliders such as the LHC, searches for top-quark FCNC interactions can be pursued through two complementary approaches, namely the study of rare top decays in top-antitop pair production events (for example from $pp \to t\bar{t} \to qX\, Wb$ where $X$ denotes a SM gauge or Higgs boson) and the investigation of FCNC-induced single top production processes (like $pp \to tX$). In this work, we consider the joint effect of these two channels as probes to non-standard top-quark couplings to lighter quarks and neutral bosons. In addition, to interpret potential deviations from the SM predictions in a consistent and model-independent way, we adopt the SMEFT framework~\cite{Buchmuller:1985jz, Grzadkowski:2010es, Brivio:2017vri} where the SM Lagrangian ${\cal{L}}_\mathrm{SM}$ is extended by higher-dimensional operators constructed from the SM fields and respecting the full $SU(3)_C \times SU(2)_L \times U(1)_Y$ gauge symmetry,
\begin{equation}
  \mathcal{L}_\mathrm{EFT} = \mathcal{L}_\mathrm{SM} + \sum_{\mathcal{D} \geq 5} \frac{C_i \, \mathcal{O}_i^{(\mathcal{D})}}{\Lambda^{\mathcal{D}-4}}\,.
\end{equation}
Here $\mathcal{O}_i^{(\mathcal{D})}$ represent gauge-invariant operators of mass dimension $\mathcal{D}$, $C_i$ are the associated dimensionless Wilson coefficients encoding the strength of the corresponding interactions and $\Lambda$ denotes the energy scale of new physics. In general, the leading new-physics effects are then assumed to be captured by the subset of dimension-six operators, which provides a controlled and predictive expansion valid in a regime where the typical scale inherent to the considered phenomena is smaller than~$\Lambda$.

In the case of top-quark FCNC interactions, several dimension-six SMEFT operators contribute to both top decay and production processes. These include operators modifying the top's interactions with the SM gauge bosons and light quarks, as well as four-fermion interactions involving at least one top-quark field. The relevant operators for this work are of the first category, and they can be grouped, following the normalisation convention and naming scheme of \cite{Zhang:2014rja}, as
\begin{equation}\begin{split}
  \mathcal{O}_{\varphi q}^{1(ij)} = \frac{y_t^2}{2}\, (\varphi^{\dagger} i \overleftrightarrow{D}_{\!\!\mu} \varphi)\, (\bar{q}_i \gamma^{\mu} q_j)\,, \qquad\qquad
  & \mathcal{O}_{\varphi q}^{3(ij)} = \frac{y_t^2}{2}\, (\varphi^{\dagger} i \overleftrightarrow{D}_{\!\!\mu}^{\!I}\varphi)\,(\bar{q}_i \gamma^{\mu} \tau^I q_j)\,, \\
  \mathcal{O}_{\varphi u}^{(ij)} = \frac{y_t^2}{2}\, (\varphi^{\dagger} i \overleftrightarrow{D_{\mu}} \varphi)\,(\bar{u}_i \gamma^{\mu} u_j)\,, \qquad\qquad
  & \mathcal{O}_{u\varphi}^{(ij)} = -y_t^3\, (\bar{q}_i u_j \tilde{\varphi})\, (\varphi^{\dagger} \varphi)\,, \\
  \mathcal{O}_{uW}^{(ij)} = y_t g\, (\bar{q}_i \sigma^{\mu\nu} \tau^I u_j)\, \tilde{\varphi} W_{\mu\nu}^I\,, \qquad\qquad
  & \mathcal{O}_{uB}^{(ij)} = y_t g_Y\, (\bar{q}_i \sigma^{\mu\nu} u_j)\, \tilde{\varphi} B_{\mu\nu}\,, \\[.15cm]
  \mathcal{O}_{uG}^{(ij)} = y_t g_s\, (\bar{q}_i \sigma^{\mu\nu} T^A u_j)\, \tilde{\varphi} G_{\mu\nu}^A\,.\qquad\qquad &
\end{split}\label{eq:smeft}\end{equation}
In these expressions, $q_i$ and $u_i$ denote the left-handed quark doublet and right-handed up-type quark singlet fields of generation $i$, respectively, while $\varphi$ represents the SM Higgs doublet (with $\tilde{\varphi} = i \tau^2 \varphi^*$). Moreover, $B_{\mu\nu}$, $W_{\mu\nu}^I$ and $G_{\mu\nu}^A$ are the field strength tensors of the $U(1)_Y$, $SU(2)_L$ and $SU(3)_C$ gauge fields, with the corresponding gauge couplings being given by $g_Y$, $g$ and $g_s$. In addition, we denote by $\tau^I$ and $T^A$ the $SU(2)$ and $SU(3)$ generators in the fundamental representation, respectively, and by $Y$ the hypercharge operator. Finally, the covariant derivative operator acting on the Higgs field is defined as usual, with $D_\mu = \partial_\mu - i g \tau^I W_\mu^I - i g_Y Y B_\mu$, and the bidirectional derivatives are given by
\begin{equation}
\varphi^\dagger i \overleftrightarrow{D}_{\!\!\mu} \varphi \equiv i \varphi^\dagger (D_\mu \varphi) - i (D_\mu \varphi)^\dagger \varphi\,, \qquad
\varphi^\dagger i \overleftrightarrow{D}_{\!\!\mu}^{\!I} \varphi \equiv i \varphi^\dagger \tau^I (D_\mu \varphi) - i (D_\mu \varphi)^\dagger \tau^I \varphi\,.
\end{equation}
We also emphasise that according to the normalisation convention of~\cite{Zhang:2014rja}, explicit powers of the top Yukawa coupling $y_t$ are factored out in the operator definitions. Among the electroweak operators $\mathcal{O}_{\varphi q}^{1}$ and $\mathcal{O}_{\varphi q}^{3}$, only the linear combination $\mathcal{O}_{\varphi q}^{-(ij)} \equiv \mathcal{O}_{\varphi q}^{1(ij)} - \mathcal{O}_{\varphi q}^{3(ij)}$ contributes to up-sector FCNC transitions at tree level, with the orthogonal combination $\mathcal{O}_{\varphi q}^{+}$ defined in a similar way affecting instead down-type quark interactions. Restricting our analysis to flavour-changing interactions between the top quark and the first two generations of light up-type quarks, we therefore define a BSM parameter space built on a set of ten independent complex parameters per light quark flavour $a=1, 2$~\cite{Durieux:2014xla},
\begin{equation}\begin{split}
  C_{\varphi q}^{-(a3)} = C_{\varphi q}^{-(3a)*} \equiv C_{\varphi q}^{-(a+3)}\,,\qquad
  C_{\varphi u}^{ (a3)} = C_{\varphi u}^{ (3a)*} \equiv C_{\varphi u}^{ (a+3)}\,,\qquad 
  C_{u\varphi}^{(a3)}\,,\qquad
  C_{u\varphi}^{(3a)}\,,\\[.1cm]
  C_{uB}^{(a3)}\,,\qquad 
  C_{uB}^{(3a)}\,,\qquad
  C_{uW}^{(a3)}\,,\qquad
  C_{uW}^{(3a)}\,, \qquad
  C_{uG}^{(a3)}\,,\qquad
  C_{uG}^{(3a)}\,.
\end{split}\end{equation}

After electroweak symmetry breaking, the SMEFT operators above induce effective FCNC interactions between the top quark and the SM gauge bosons. For practical purposes in collider analyses such as the one presented in this study, it is convenient to work directly in the broken phase where fields are expressed in terms of the physical mass eigenstates. The relevant interactions can then be captured by an effective Lagrangian in which each coupling depends on one or more SMEFT Wilson coefficients, 
\begin{equation}\begin{split}
  \mathcal{L}_\text{eff} =\ & 
    - \frac{g}{2 c_W}\, \bar{t} \gamma^{\mu} (v_{tq}^{Z} - a_{tq}^{Z} \gamma_5)\, q\, Z_{\mu}
    - \frac{g}{2\sqrt{2}}\, \bar{q} (g^v_{qt} + g^a_{qt} \gamma_5)\, t\, h \\
  & - \frac{e}{\Lambda}\, \bar{t} \sigma^{\mu\nu} (f_{tq}^{\gamma} + i h_{tq}^{\gamma} \gamma_5)\, q\, A_{\mu\nu}
    - \frac{g}{2 c_W \Lambda}\, \bar{t} \sigma^{\mu\nu} (f_{tq}^{Z} + i h_{tq}^{Z} \gamma_5)\, q\, Z_{\mu\nu} \\
  & - \frac{g_s}{\Lambda}\, \bar{t} \sigma^{\mu\nu} T^A (f_{tq}^{g} + i h_{tq}^{g} \gamma_5)\, q\, G_{\mu\nu}^A 
    + \text{H.c.}
\end{split}\label{effLag}\end{equation}
In this Lagrangian, $q = u, c$ denotes again a light up-type quark, while the effective couplings $v_{tq}^{Z}$, $a_{tq}^{Z}$, $g_{qt}^{v,a}$, $f_{tq}^{V}$ and $h_{tq}^{V}$ (with $V = \gamma, Z, g$) parametrise the strength of the FCNC interactions of the top quark with the Higgs boson $h$, the $Z$ boson, the photon $A$ and the gluon $g$, respectively, with the corresponding gauge boson field-strength tensors being represented by $A_{\mu\nu}$, $Z_{\mu\nu}$ and $G_{\mu\nu}^A$. Moreover, we conventionally introduced in the prefactors of the top FCNC couplings to the $Z$ boson a dependence on the cosine of the electroweak mixing angle $c_W$. This broken-phase parametrisation features only two dimension-four operators (the vector and scalar currents) and three dimension-five dipole operators per flavour. While it does not reflect the full gauge structure of SMEFT at high energies, it provides, as said above, a convenient and widely used framework for modelling FCNC processes in the top-quark sector~\cite{Aguilar-Saavedra:2004mfd, Aguilar-Saavedra:2008nuh, Durieux:2014xla}.

In the present work, we quantitatively explore the impact of deep learning methods on probing the properties of the top quark. As a proof of concept, we focus on top FCNC interactions involving a photon and consider both the rare top decay $t \to q\gamma$ that is considered to follow the production of a top-antitop pair and the associated production process $qg \to t\gamma$. In this case, the photonic dipole couplings $f_{tq}^\gamma$ and $h_{tq}^\gamma$ arise from a linear combination of the dimension-six operators of Eq.~\eqref{eq:smeft} involving the hypercharge and weak gauge field strengths,
\begin{equation}\begin{split}
  -\frac{e}{\Lambda} f^\gamma_{tq} = e \frac{m_t}{\Lambda^2} \bigg[ \Big(C_{uB}^{(3a)} + C_{uW}^{(3a)}\Big) + \Big(C_{uB}^{(a3)} + C_{uW}^{(a3)}\Big)^* \bigg] \,,\\[.2cm]
  -i \frac{e}{\Lambda} h^\gamma_{tq} = e \frac{m_t}{\Lambda^2} \bigg[ \Big(C_{uB}^{(3a)} + C_{uW}^{(3a)}\Big) - \Big(C_{uB}^{(a3)} + C_{uW}^{(a3)}\Big)^* \bigg] \,.
  \label{eq:WCs:params}
\end{split}\end{equation}
We will then assess the sensitivity of analysis strategies relying on either a traditional cut-and-count approach or deep learning methods, and to interpret the results in terms of bounds on the effective dipole couplings $f_{tq}^\gamma$ and $h_{tq}^\gamma$ introduced in Eq.~\eqref{effLag}. However, experimental constraints on top FCNC interactions are commonly reported in terms of upper bounds on the top rare branching ratios. We therefore provide expressions for the partial width of the decay $t \to q \gamma$ in terms of the dipole coefficients~\cite{Beneke:2000hk, Aguilar-Saavedra:2004mfd},
\begin{equation}
\Gamma(t \to q\gamma) = \frac{2 \alpha}{\Lambda^2}\, \Big( |f_{tq}^\gamma|^2 + |h_{tq}^\gamma|^2 \Big) \, m_t^3 \,,
\label{eq:t_to_gammaq}\end{equation}
where $\alpha=e^2/(4\pi)$ is the electromagnetic coupling constant and $m_t$ is the top-quark mass. To extract the associated branching ratio, we normalise this result to the total width of the top quark that includes, in addition to Eq.~\eqref{eq:t_to_gammaq}, the dominant contribution from the SM process $t \to bW^+$. At next-to-leading order in the strong coupling $\alpha_s = g_s^2/(4\pi)$, the latter is given by~\cite{Beneke:2000hk}
\begin{equation}
  \Gamma(t \to bW^+) = \frac{G_F m_t^3}{8\pi \sqrt{2}} \left( 1 - \frac{m_W^2}{m_t^2} \right)^2 \left( 1 + 2 \frac{m_W^2}{m_t^2} \right)  \left[ 1 - \frac{2\alpha_s}{3\pi} \left( \frac{2\pi^2}{3} - \frac{5}{2} \right) \right] \,,
\end{equation}
which yields $\Gamma(t \to bW^+) = 1.329$ GeV for a top mass of $m_t = 172.5$~GeV, a $W$-boson mass of $m_W = 80.419$~GeV and a Fermi constant fixed to $G_F = 1.16637 \times 10^{-5}$~GeV$^{-2}$. The branching ratio for the photonic FCNC top decay $t\to q\gamma$ can then be written in a semi-analytical form as
\begin{equation}
  \mathrm{BR}(t \to q\gamma) \approx 0.060\, \Big( |f_{tq}^\gamma|^2 + |h_{tq}^\gamma|^2 \Big) \,,
\label{eq:br_tqgamma}\end{equation}
which will serve as the basis for interpreting the sensitivity of our collider analyses to the anomalous top–photon couplings.

Over the past decade, both the CMS and ATLAS collaborations have conducted dedicated searches for FCNC interactions involving the top quark and a photon. The first constraints on the $\mathrm{BR}(t \to q\gamma)$ branching ratio of Eq.~\eqref{eq:br_tqgamma} came from CMS, using the full 8~TeV dataset to set limits of BR$(t \rightarrow u\gamma) < 1.3 \times 10^{-4}$ and BR$(t \rightarrow c\gamma) < 1.7 \times 10^{-3}$ at 95\% confidence level (CL)~\cite{CMS:2015kek}. More recently, ATLAS improved these bounds using 139~fb$^{-1}$ of 13~TeV data, reporting BR$(t \rightarrow u\gamma) < (0.85-1.2) \times 10^{-5}$ and BR$(t \rightarrow c\gamma) < (4.2-4.5) \times 10^{-5}$ at 95\% CL~\cite{ATLAS:2022per}, the precise threshold depending on the Lorentz structure of the top FCNC interactions. However, the latest and most stringent constraints have been established by CMS using the full Run 2 dataset~\cite{CMS:2023bjm}. This last analysis combined events originating from single top production with an associated photon with $t\bar{t}$ events featuring a rare top decay. It then employed multivariate techniques to efficiently suppress the SM background, obtaining 95\% CL upper limits of BR$(t \rightarrow \gamma u) < 0.95 \times 10^{-5}$ and BR$(t \rightarrow \gamma c) < 1.51 \times 10^{-5}$.

Looking ahead, the HL-LHC is expected to significantly improve the sensitivity to rare top decays. Operating at $\sqrt{s} = 14$~TeV and aiming for an integrated luminosity of 3000~fb$^{-1}$, projections suggest that branching fractions as low as $10^{-6}$ could be probed~\cite{Mandrik:2018gud}. Additionally, even more stringent limits are anticipated from next-generation colliders. A future circular hadron collider foreseen to operate at a centre-of-mass energy of $\sqrt{s} = 100$~TeV with 30~ab$^{-1}$ of data could hence push the sensitivity down to BR$(t \rightarrow u\gamma) < 1.8 \times 10^{-7}$ and BR$(t \rightarrow c\gamma) < 2.4 \times 10^{-7}$~\cite{Mandrik:2018yhe}, while other proposed facilities such as TESLA~\cite{Aguilar-Saavedra:2001ajk} and future muon colliders~\cite{Ake:2023xcz} could in principle provide complementary information due to their clean experimental environments and potential for polarisation control.

%%%%%%%%%%%%%%%%%%%%%%%%%%%%%%%%%%%%%%%%%%%%%
\section{Signal and backgrounds}
\label{S:topFCNC}
%%%%%%%%%%%%%%%%%%%%%%%%%%%%%%%%%%%%%%%%%%%%%

%%%%%%%%%%%%%%%%%%%%%%%%%%%%%%%%%%%%%%%%%%%%%%%%%%%%%%%%%%%%%%
\subsection{Modelling signal processes at the LHC}\label{sec:modelling}
%%%%%%%%%%%%%%%%%%%%%%%%%%%%%%%%%%%%%%%%%%%%%%%%%%%%%%%%%%%%%%

\begin{figure}
  \centering \includegraphics[width=0.75\linewidth]{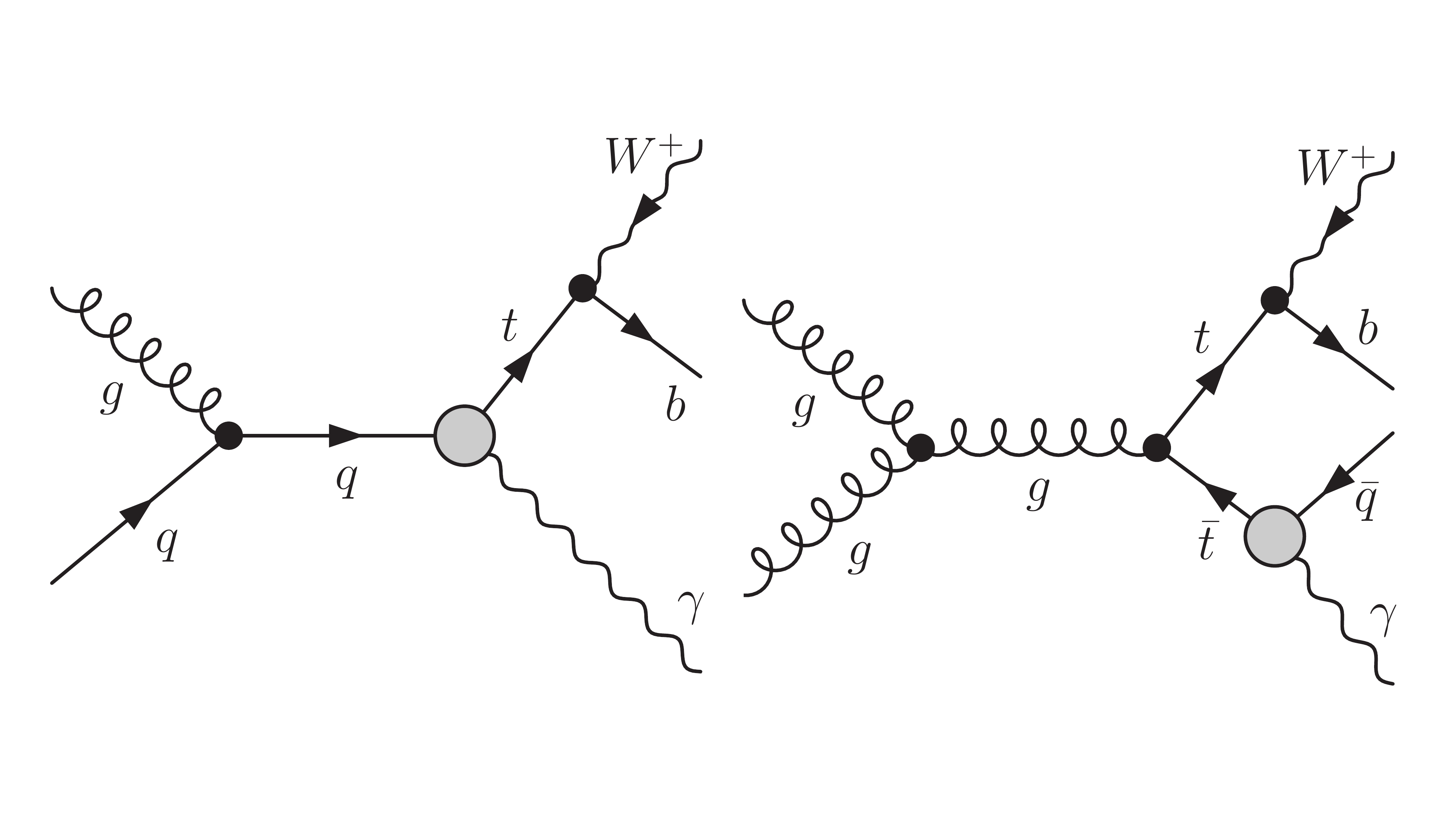} \vspace{-0.4cm}
  \caption{Representative tree-level Feynman diagrams for the FCNC production of a single SM-decaying top quark in association with a photon (left), and for the SM production of a top-antitop pair where one the tops decays in the SM way and the other through an FCNC process (right). \label{fig:FD:tqy}}
\end{figure}

As discussed in section~\ref{S:model}, the FCNC couplings of the top quark to a photon and either an up quark or a charm quark can be investigated at hadron colliders through two main channels. The first involves the production of a top quark in association with a photon, with the top subsequently decaying into a bottom quark and a $W$ boson. The second channel probes FCNC interactions in SM-induced top pair production where one top decays through the SM mode $t\to b W$ and the other via an FCNC process $t\to \gamma q$. Representative tree-level Feynman diagrams for the two processes are shown in figure~\ref{fig:FD:tqy}.

Each production mode presents distinct advantages and challenges. In single top plus photon production, the photon typically recoils against the top quark. This leads to event final states comprising a photon with large transverse momentum, such a feature helping discriminate signal from SM backgrounds. However, disentangling the properties of the top-photon FCNC couplings in this channel is more complex as all four couplings appearing in Eq.~\eqref{effLag}, namely $f_{tu}^\gamma$, $f_{tc}^\gamma$, $h_{tu}^\gamma$ and $h_{tc}^\gamma$, contribute in different combinations modulated by the corresponding parton distribution functions. In contrast, the top pair production mode benefits from a simpler structure: the new physics amplitude factorises cleanly with respect to the FCNC couplings. Yet, this channel presents its own difficulty as the final-state kinematics closely resemble those of the irreducible SM background, complicating signal extraction as will be discussed in the following sections.

\begin{figure}
    \centering
    \includegraphics[width=0.32\linewidth]{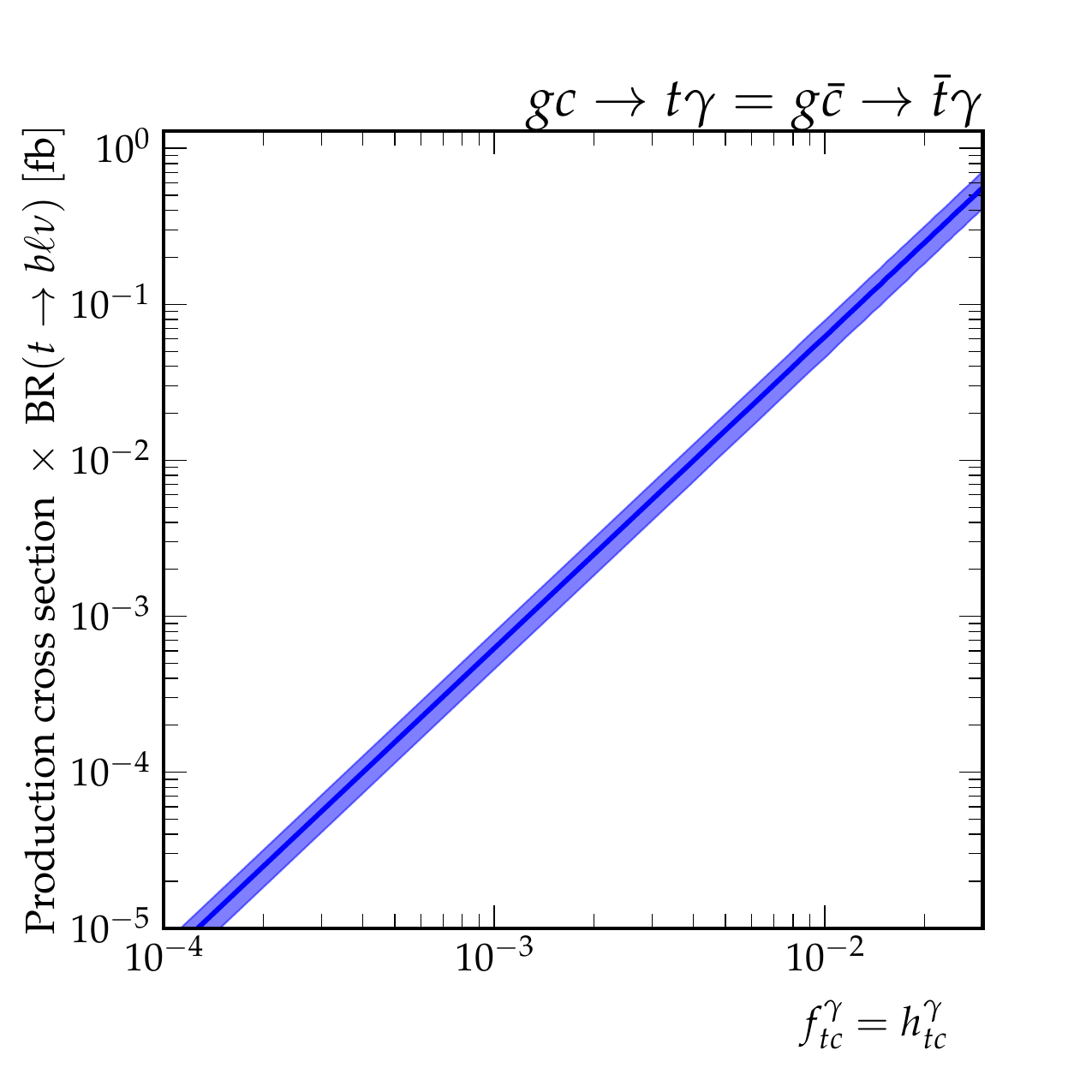}
    \includegraphics[width=0.32\linewidth]{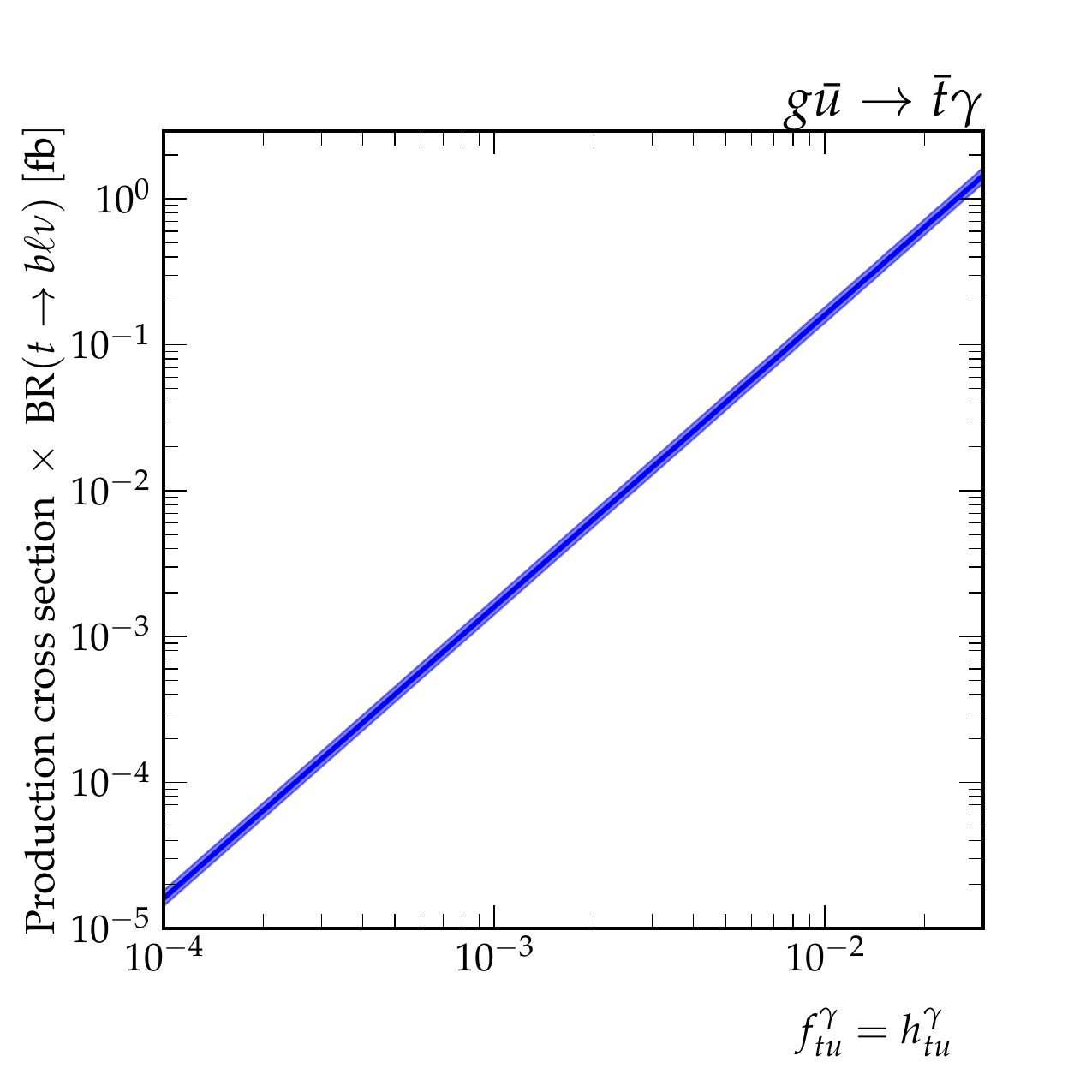}
    \includegraphics[width=0.32\linewidth]{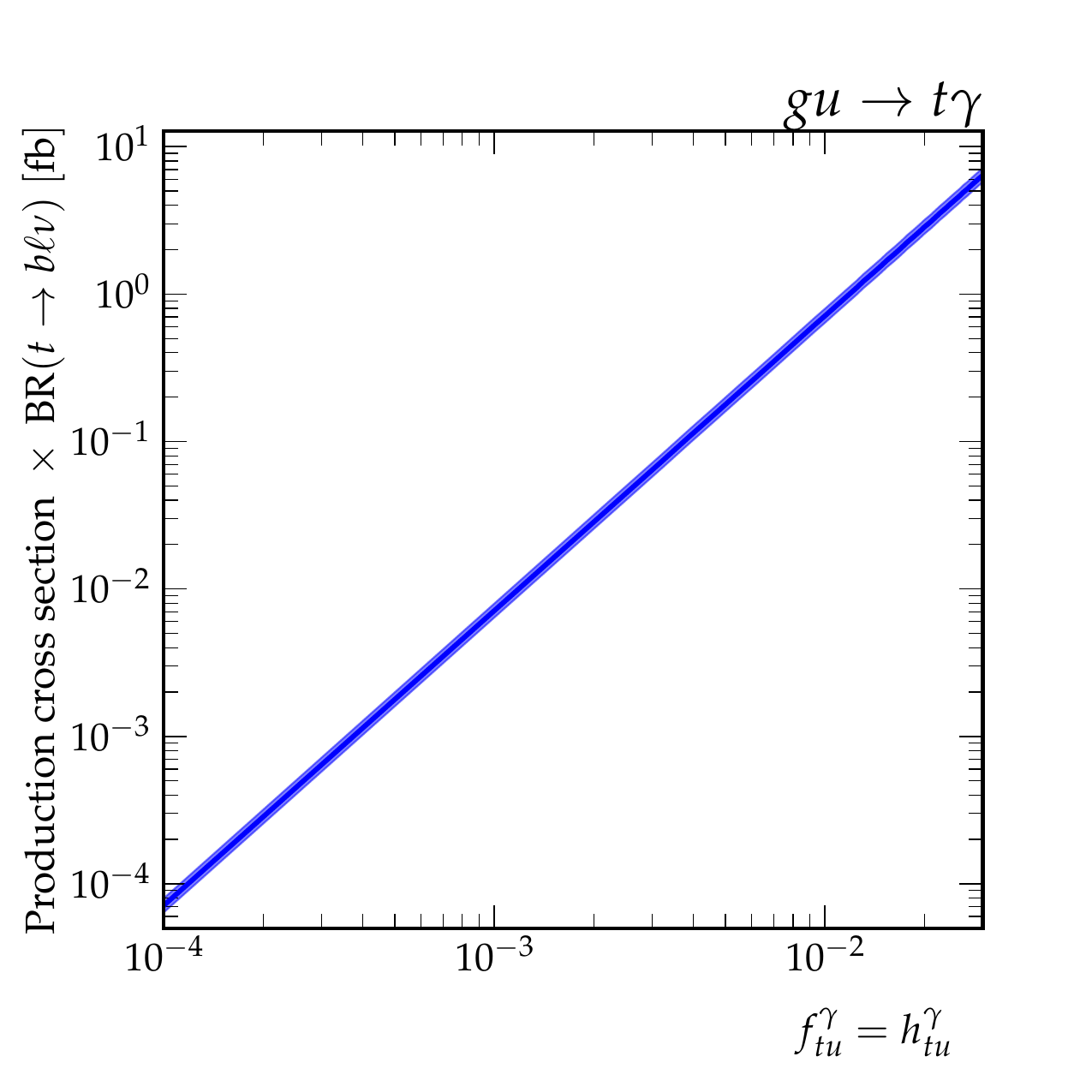}
    \caption{Leading-order cross sections for single top production in association with a photon via $cg$ (or equivalently $\bar{c}g$) fusion (left), $\bar{u}g$ fusion (middle) and $ug$ fusion (right). The bands indicate theoretical uncertainties from scale and PDF variations summed in quadrature.\label{fig:xsec:SP}}
\end{figure}

We now turn to the expected signal cross sections for both the single and pair production channels. We begin with the single top quark production process in association with a photon, $pp \to t\gamma$, where $t$ generically denotes both top and antitop quarks. The total hadronic cross section at leading order can be expressed as
\begin{equation}
    \sigma(pp \to t\gamma) = \sum_{a=u,\bar{u},c,\bar{c}} \int \mathrm{d} x_a\, \mathrm{d}x_g\, f_{a/p}(x_a; \mu_F^2)\, f_{g/p}(x_g; \mu_F^2)\, \hat{\sigma}_{ag \to t\gamma}(\hat{s}; \mu_F^2, \mu_R^2)\,,
\label{eq:xsec} \end{equation}
where $f_{i/p}(x; \mu_F^2)$ denotes the parton distribution function (PDF) of parton $i$ in the proton, evaluated at momentum fraction $x$ and factorisation scale $\mu_F$. Moreover, in this expression, the partonic cross section $\hat{\sigma}$ is integrated over the two-body phase space and thus only depends on the partonic centre-of-mass energy $\hat{s} = x_a x_g s$, with $s$ being the hadronic collision energy, and the unphysical scales $\mu_F$ and $\mu_R$.  At tree level, two subprocesses contribute: an $s$-channel diagram where the incoming (anti)quark couples to the gluon (as shown in the left diagram of figure~\ref{fig:FD:tqy}), and a diagram featuring the $t$-channel exchange of a top or antitop quark. Using the effective Lagrangian of Eq.~\eqref{effLag}, the corresponding partonic cross section scales as
\begin{equation}
    \hat{\sigma}_{qg \to t\gamma} \propto \frac{g_s^2 e^2}{\pi \Lambda^2 \hat{s}^2} \bigg(|f_{tq}^\gamma|^2 + |h_{tq}^\gamma|^2 \bigg)\,,
\label{eq:xsec:partonic}\end{equation}
neglecting constant numerical and phase space prefactors irrelevant to our present discussion. The relative importance of the different contributions is thus primarily encoded in the associated parton luminosities appearing in Eq.~\eqref{eq:xsec}. Assuming the simplifying scenario in which the couplings satisfy $f_{tu}^\gamma = f_{tc}^\gamma$ and $h_{tu}^\gamma = h_{tc}^\gamma$, the ordering of the cross sections for the different initial-state flavours is given by
\begin{equation}
    \sigma(ug \to t\gamma) > \sigma(\bar{u}g \to \bar{t}\gamma) > \sigma(\bar{c}g \to \bar{t}\gamma) \approx \sigma(cg \to t\gamma)\,,
\end{equation}
as illustrated in figure~\ref{fig:xsec:SP} for a hadronic centre-of-mass energy $\sqrt{s}=13.6$~TeV. These predictions are obtained using \textsf{Madgraph\_aMC@NLO} (v3.4.1)~\cite{Alwall:2014hca}, with the \textsf{NNPDF\_40\_lo\_as\_01180} PDF set~\cite{NNPDF:2021njg} and renormalisation and factorisation scales set to the average transverse mass of the final-state particles, and an implementation of the relevant FCNC operators in \textsf{FeynRules}~\cite{Christensen:2009jx, Alloul:2013bka} that has been exported in the \textsf{UFO} format~\cite{Degrande:2011ua, Darme:2023jdn}. The resulting cross sections are found to span several orders of magnitude, from 1--10~fb for $\mathcal{O}(1)$ couplings down to 0.01--0.1~ab for couplings around $10^{-4}$ with the quadratic FCNC coupling dependence given in Eq.~\eqref{eq:xsec:partonic}.

The analysis of the pair production channel is more straightforward, as the FCNC interaction only appears in the decay of one of the top quarks. Under the narrow-width approximation, the cross section for this FCNC signal is factorised as
\begin{equation}
    \sigma(pp \to t\bar{t} \to q\gamma\, b W) = 2 \times \sigma(pp \to t\bar{t}) \times \text{BR}(t \to q\gamma) \times \text{BR}(t \to bW)\,.
\end{equation}
This form for the cross section illustrates a clean separation between the QCD-driven $t\bar t$ production mechanism, one FCNC-induced top decay and one SM top decay. We compute the inclusive $t\bar{t}$ production cross section using \textsf{Top++} v2.0~\cite{Czakon:2011xx, Czakon:2013goa}, including  next-to-next-to-leading-order (NNLO) QCD corrections matched to next-to-next-to-leading-logarithmic (NNLL) soft-gluon resummation. We use this time the \textsf{NNPDF\_40\_nnlo\_as\_01180} PDF set, and we fix both the renormalisation and factorisation scales to the top mass.

\begin{figure}
    \centering
    \includegraphics[width=0.49\linewidth]{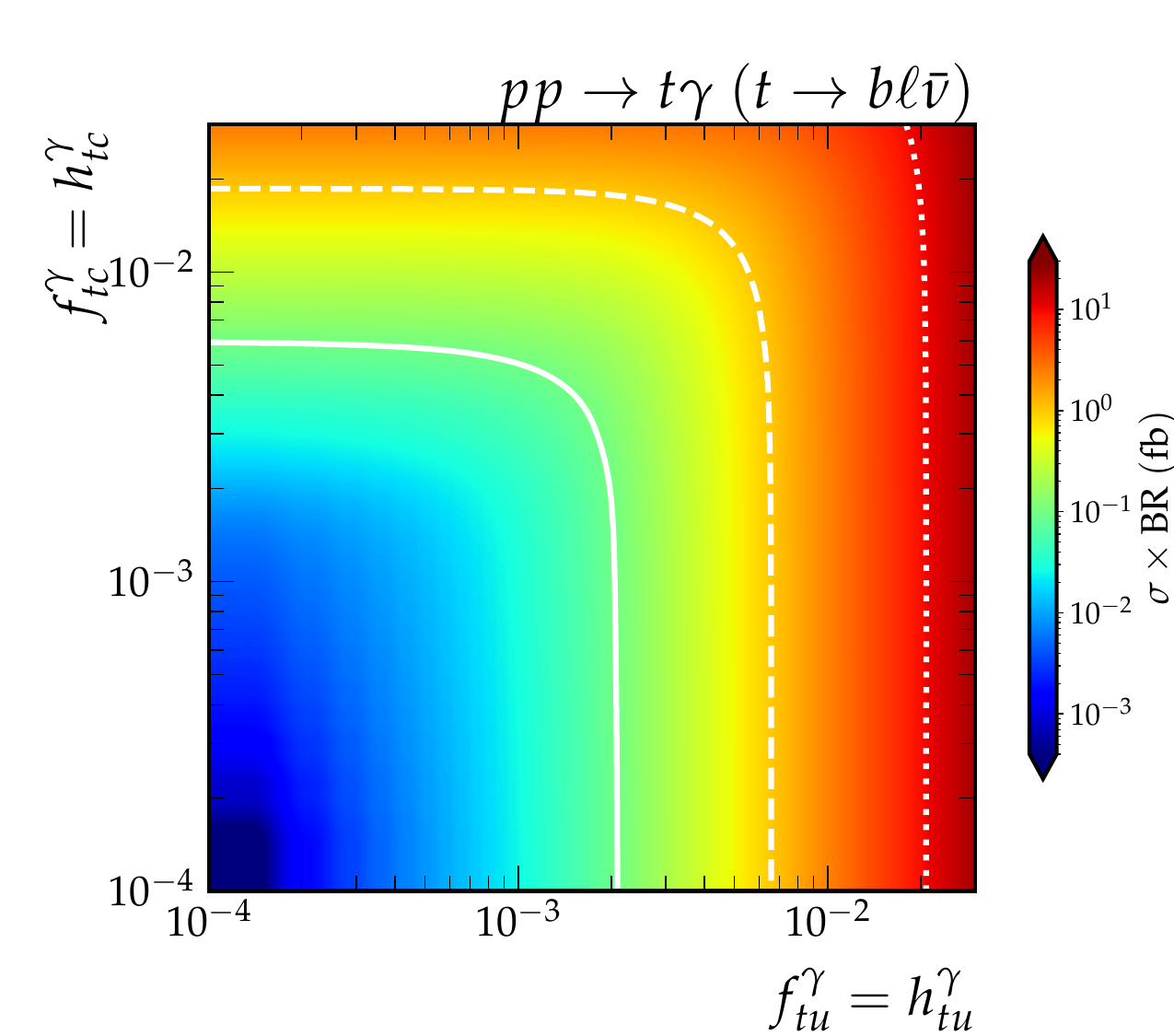}
    \includegraphics[width=0.49\linewidth]{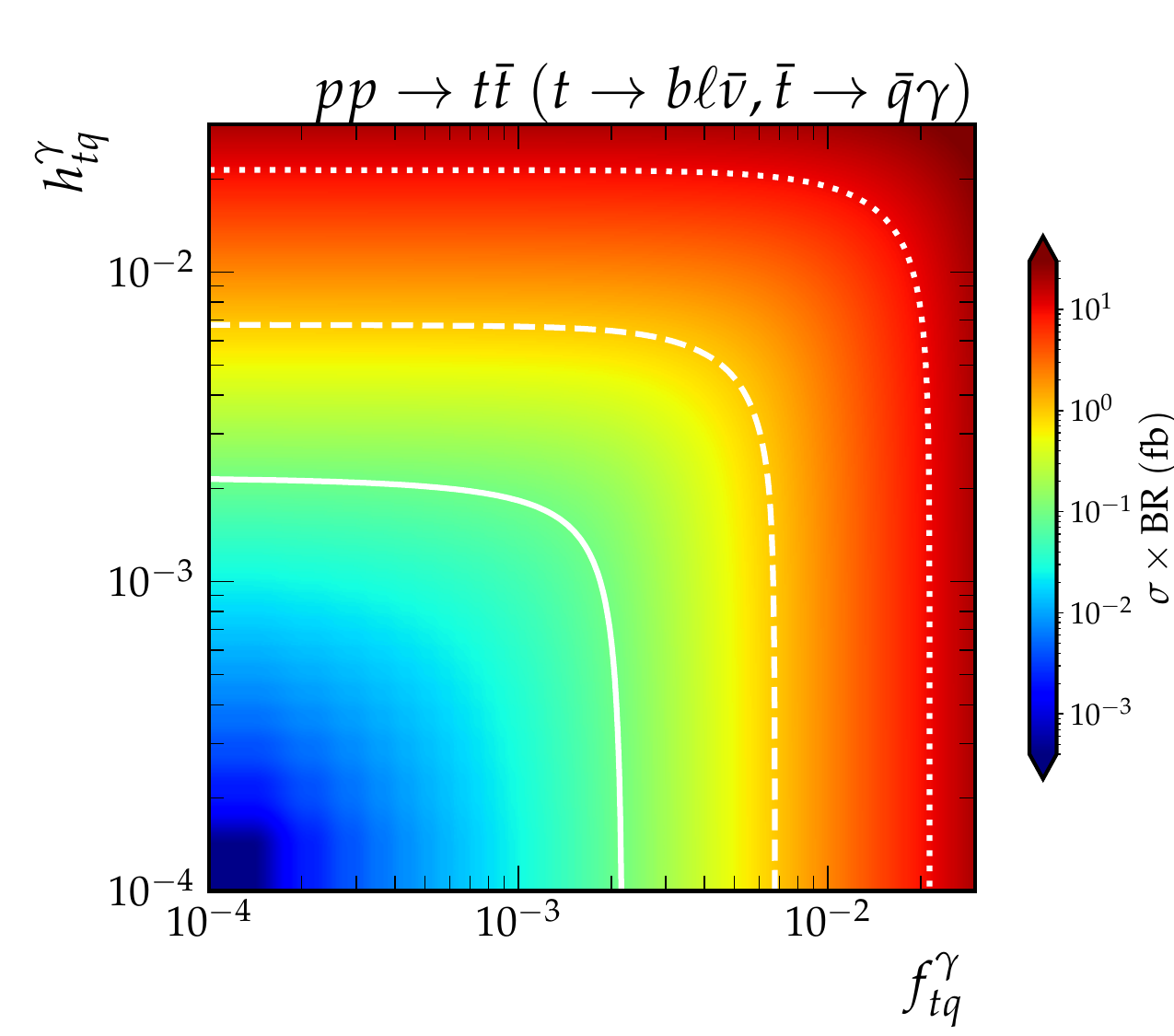}
    \caption{Signal rates $\sigma \times \text{BR}$ for single top + photon production (left) and top pair production with an FCNC decay (right). The dotted, dashed and solid contours correspond to constant cross section values of 0.1, 1 and 10~fb respectively.\label{fig:Xsec}}
\end{figure}

A comparative view of the expected signal rates at $\sqrt{s}=13.6$~TeV is presented in figure~\ref{fig:Xsec}, showing $\sigma \times \text{BR}$ for the single (left panel) and pair (right panel) production modes. For single production, we include both top and antitop contributions and normalise the predictions to include the leptonic branching ratio of the top quark that is relevant for the analysis considered in this work,
\begin{equation}
    \sigma \times {\rm BR} \equiv \bigg[\sigma(pp \to t\gamma) + \sigma(pp \to \bar{t} \gamma)\bigg] \times {\rm BR}(t\to b\ell\nu_\ell)\,,
\end{equation}
the results shown in the figure assuming $f_{tu}^\gamma = h_{tu}^\gamma$ and $f_{tc}^\gamma = h_{tc}^\gamma$. However, due to the dominance of valence up-quark initiated processes, the rate is largely dominated by contributions from the $tu\gamma$ coupling. With this setup, cross sections of $\mathcal{O}(0.1)$~fb can be achieved for couplings as small as $10^{-3}$, and reach up to 20~fb for $f_{tu}^\gamma \sim 2 \times 10^{-2}$. In contrast, the pair production mode is equivalently sensitive to both the $tu\gamma$ and $tc\gamma$ couplings depending on the considered top FCNC decay, 
\begin{equation}
    \sigma \times {\rm BR} \equiv 2 \times \sigma(pp \to \bar{t}t) \times {\rm BR}(t\to q\gamma) \times {\rm BR}(t \to b\ell\nu_\ell)\,,
\end{equation}
which yields slightly higher cross sections for the same coupling values due to the larger total production rate.

Finally, we estimate the theoretical uncertainties on the predicted signal rates by varying the renormalisation and factorisation scales independently by a factor of two around their central values, as well as by using the PDF error sets provided with the NNPDF~4.0 fits. In the single production channel estimated at LO, these uncertainties are typically of 15--25\%, dominated by PDF variations for charm-initiated processes and by scale dependence in the up-quark channel. In the pair production mode, the uncertainties drop below 10\% thanks to the precise knowledge of the inclusive $t\bar{t}$ cross section at NNLO+NNLL.

%%%%%%%%%%%%%%%%%%%%%%%%%%%%%%%%%%%%%%%%%%%
\subsection{Technical setup}
%%%%%%%%%%%%%%%%%%%%%%%%%%%%%%%%%%%%%%%%%%%

\begin{table}
\setlength\tabcolsep{11pt}\renewcommand{\arraystretch}{1.2}
\begin{center}
\begin{tabular}{l ccc}
Process & $\sigma_{\rm LO}$~[fb] & $N_{\rm events}$ & $\omega = \sigma_{\rm LO} / N_{\rm events}$ [fb] \\
\toprule
$p p \to t ~\bar{t}~\gamma$  & $5.14 \times 10^2$ & $11 \times 10^6$ & $4.67 \times 10^{-5}$ \\
$p p \to t ~j~\gamma$ & $4.14 \times 10^2$ & $8 \times 10^6$ & $5.17 \times 10^{-5}$ \\
$p p \to t ~W^\pm~\gamma$ & $5.08 \times 10^1$ & $8 \times 10^6$ & $7.10 \times 10^{-6}$  \\
$p p \to t ~b~\gamma$  & $5.62 \times 10^0$ & $8 \times 10^6$  & $7.02 \times 10^{-7}$\\
$p p \to W^\pm~(\to \ell^\pm \nu) \gamma + {\rm jets}$ & $3.77 \times 10^{4}$ & $3.29 \times 10^7$ & $1.14 \times 10^{-3}$ \\
$p p \to Z/\gamma^*~(\to \ell^+ \ell^-) \gamma + {\rm jets}$ & $2.22 \times 10^4$ & $1.5 \times 10^7$ & $1.48 \times 10^{-3}$ \\
\toprule
Signal~($p p \to t~\gamma$) & $2.87 \times 10^1$ & $8 \times 10^6$ & $2.91 \times 10^{-6}$ \\
Signal~($p p \to t~\bar{t}$) & $3.94 \times 10^1$ & $8 \times 10^6$ & $4.92 \times 10^{-6}$ \\
\toprule
\end{tabular}
\end{center}
  \caption{Total cross sections at $\sqrt{s} = 13.6$~TeV for the signal and background processes considered in this analysis, computed at LO in QCD. For each process, the number of generated events $N_\mathrm{events}$ is listed alongside the corresponding per-event weight $\omega$ controlling the numerical accuracy of each background event sample. Representative signal cross sections are shown for the benchmark values $f_{tq}^\gamma = h_{tq}^\gamma = 2 \times 10^{-2}$ of the FCNC couplings.\label{tab:xs}}  
\end{table}

In this study, we examine top-photon FCNC effects in both the single top and $t\bar t$ pair production modes. As briefly sketched in section~\ref{sec:modelling}, in the single production mode the top quark is assumed to decay leptonically into a charged lepton (electron or muon), a neutrino and a bottom quark while in the pair production mode, we assume an SM leptonic decay for one of the top quarks while the other undergoes an FCNC decay into a photon and either an up or a charm quark. The signal signatures are therefore given by:
\begin{equation}\begin{split}
    pp \to t~(\to \ell^+ \bar{\nu}_\ell b)~\gamma\,, \qquad & pp \to \bar{t}~(\to \ell^- \nu_\ell \bar{b})~\gamma\,, \\
    pp \to t~(\to \ell^+ \bar{\nu}_\ell b)~\bar{t}~(\to \bar{q}~\gamma)\,, \qquad & pp \to t~(\to q~\gamma)~\bar{t}~(\to \ell^- \nu_\ell \bar{b})\,,
\end{split}\end{equation}
where $q$ stands for either an up or a charm quark. In total, we thus consider three distinct signal topologies in terms of final-state objects as the pair production mode leads to two separate contributions depending on the light quark in which the top decays through the FCNC channel.

The final-state signature common to all signal processes thus consists of exactly one charged lepton (electron or muon), a photon, at least one jet (with at least one $b$-tagged jet) and missing transverse energy. The corresponding backgrounds consequently arise from SM processes involving photons, jets, leptons and neutrinos, and these can be broadly categorised into three classes.
\begin{itemize}
  \item \textit{Top production with a photon and jets} -- This contribution to the background includes both top quark pair production in association with a photon ($pp\to t\bar{t}\gamma$) and single top production processes such as $pp\to tW\gamma$, $tj\gamma$ and $tb\gamma$. For the process $pp\to t\bar{t}\gamma$, we consider contributions where one top decays leptonically and the other hadronically. In the $tW\gamma$ channel, both leptonic and hadronic decays of the $W$ boson and the top quark are included, while for the $pp\to tj\gamma$ and $pp\to tb\gamma$ processes, the top quark is always enforced to decay leptonically.

  \item \textit{Electroweak boson production with a photon and jets} -- This contribution to the background includes $W\gamma$ and $Z\gamma$ production with up to two additional jets. In the $W\gamma$ case, we require the $W$ to decay leptonically so that the full process corresponds to $pp\to \ell^\pm \nu_\ell \gamma$ + jets. On the other hand, the $Z\gamma$ background comprises the leptonic decay of a possibly off-shell $Z$ boson, and thus corresponds to the process $pp\to \ell^+ \ell^- \gamma$ + jets. It will therefore be significantly suppressed by requiring exactly one charged lepton in the final state, as dictated by the signal topology.

  \item \textit{Fake photon and lepton backgrounds} -- These contributions to the background arise when a jet is misidentified as a photon, typically through the decay of high-$p_T$ mesons into photons, or when non-prompt leptons from hadron decays are misidentified as prompt ones. In addition, electrons may be misidentified as photons, especially in processes like $pp\to t\bar{t}$, $VV$, $Z(\to e^+ e^-)$ + jets and $tX$.
\end{itemize}

Signal and background processes involving top quarks are generated at LO using the five-flavour number scheme. For $V\gamma$+jets backgrounds, exclusive samples based on matrix elements including up to two additional partons are produced and matched to the parton shower using the MLM scheme~\cite{Mangano:2006rw, Alwall:2008qv}, in order to avoid phase-space double counting. In addition, all processes employ a dynamical scale choice, defined as the average transverse mass of all massive final-state particles, and we use the \textsf{NNPDF\_40\_lo\_as\_01180} set of parton distribution functions. In our simulation tool chain, the decays of all heavy resonances such as the top quark and electroweak gauge bosons are handled using \textsf{MadSpin}~\cite{Artoisenet:2012st} and \textsf{MadWidth}~\cite{Alwall:2014bza} to preserve spin correlations. The resulting parton-level events are then passed to \textsc{Pythia}~8.309~\cite{Bierlich:2022pfr} for parton showering, hadronisation and hadron decays.

Detector effects are simulated using the Simplified Fast Simulation (SFS) module~\cite{Araz:2020lnp, Araz:2021akd} embedded within the \textsf{MadAnalysis~5} framework~\cite{Conte:2012fm, Conte:2014zja, Conte:2018vmg}, which is also used to implement the analysis strategy and prepare inputs for the machine learning models as introduced in \cite{Cornell:2024dki, Cornell:2021gut}. In this scheme, jets are reconstructed using the anti-$k_T$ algorithm~\cite{Cacciari:2008gp} with a radius parameter $R=0.4$ as implemented in \textsf{FastJet}~3.4.1~\cite{Cacciari:2011ma}. Lepton and jet reconstruction and identification performance is taken from a validated \textsf{MadAnalysis~5} routine designed for searches involving right-handed $W$ bosons and heavy neutrinos~\cite{DVN/UMGIDL_2023, CMS:2021dzb}, that has been described in detail in~\cite{Frank:2023epx}.

Finally, for the implementation of the machine learning models on which our analysis relies, we use the \textsf{PyTorch Geometric} platform~\cite{fey2019fast} to build and train graph neural networks, while the standard \textsf{PyTorch} package~\cite{paszke2019pytorch} is employed for fully connected networks and transformers. The \textsf{Scikit-Learn} library is used to facilitate preprocessing steps and evaluation metrics, whereas we rely on the \textsf{PyTorch Lightning} software~\cite{falcon2019pytorch} to manage the training workflows including model validation. For the purpose of open science, all code and analysis pipelines are available from a public \href{https://github.com/AHamamd150/Top_FCNC_Lightning}{GitHub repository}.

%%%%%%%%%%%%%%%%%%%%%%%%%%%%%%%%%%%%%%%%%%%
%\subsection{Analysis methodology}
%%%%%%%%%%%%%%%%%%%%%%%%%%%%%%%%%%%%%%%%%%%

%%%%%%%%%%%%%%%
\section{Analysis Methods}
\label{S:methods}
%%%%%%%%%%%%%%%

\begin{figure}[!t]
\centering
\includegraphics[width=0.32\linewidth]{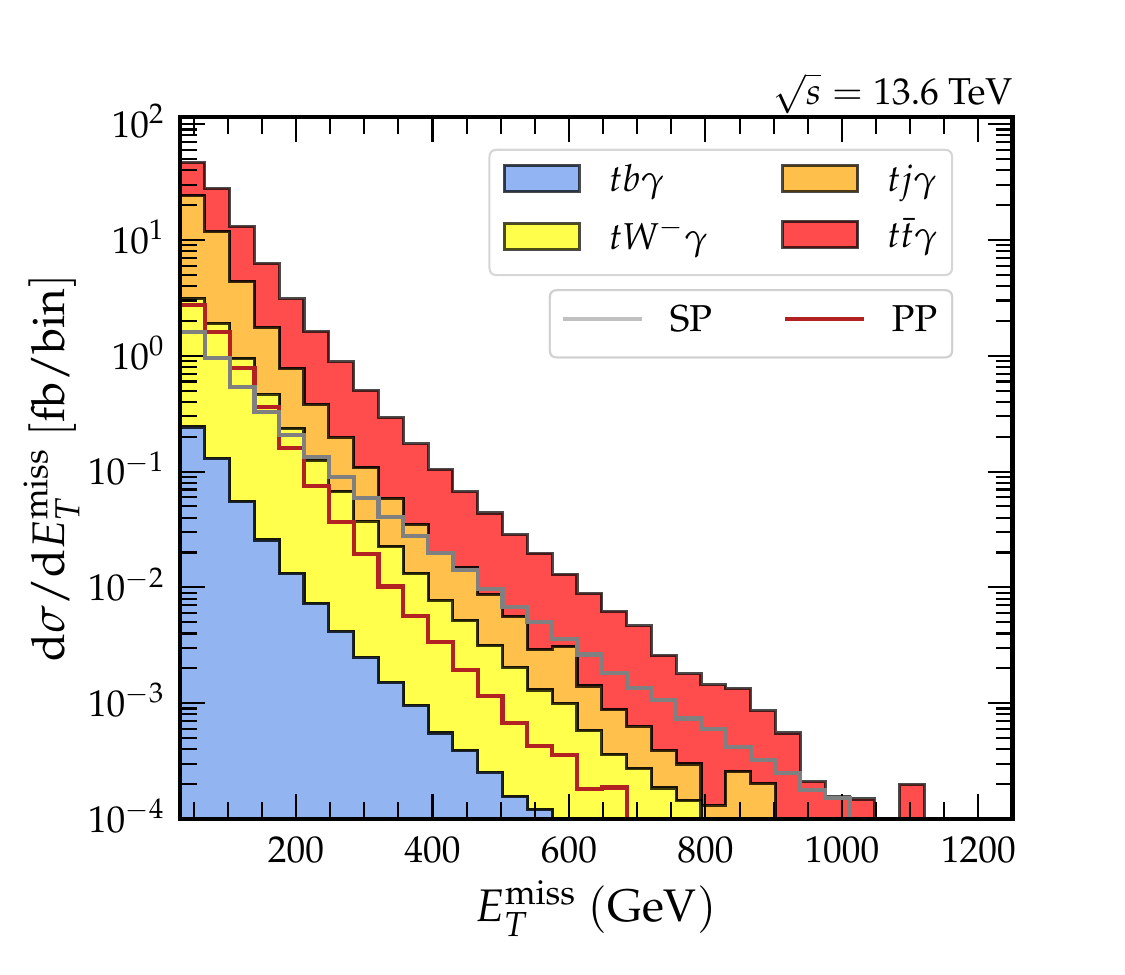}
\includegraphics[width=0.32\linewidth]{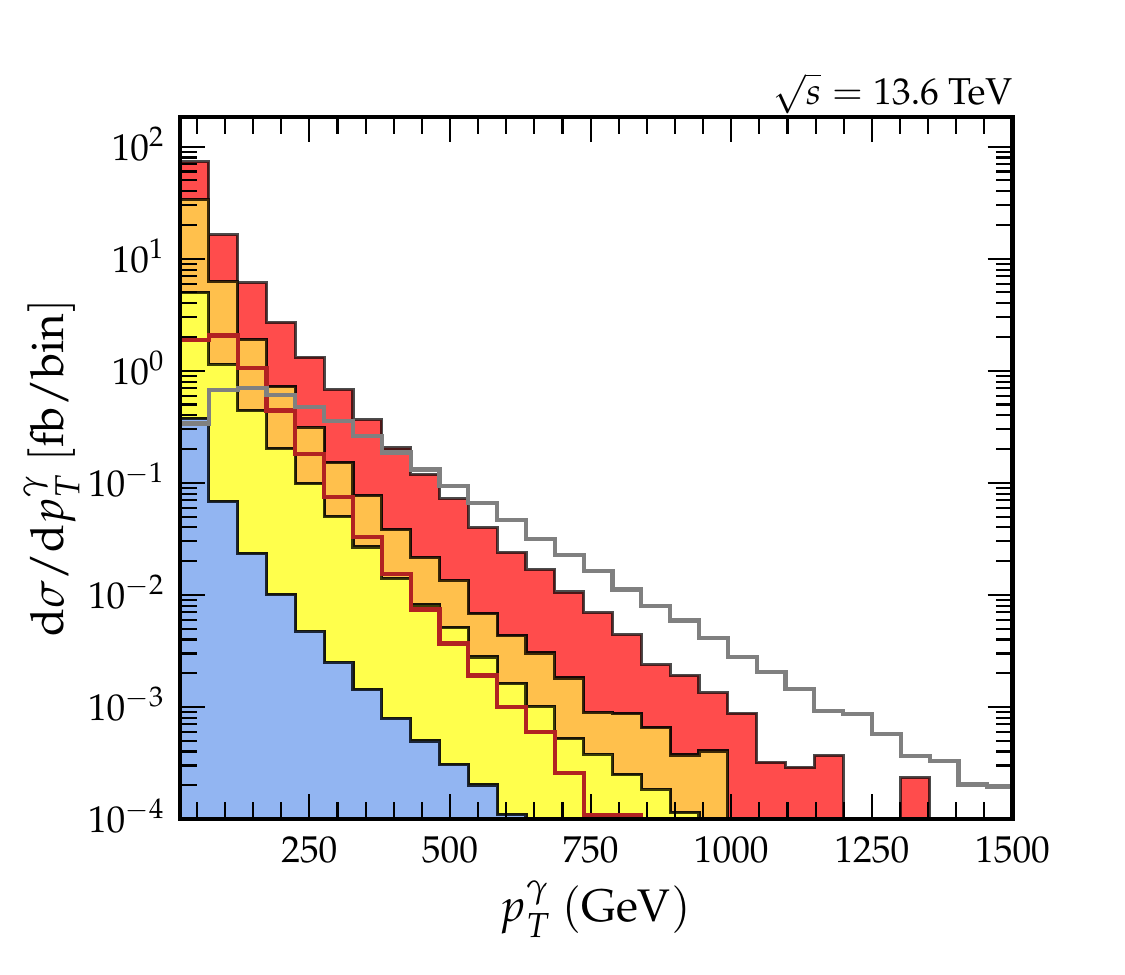}
\includegraphics[width=0.32\linewidth]{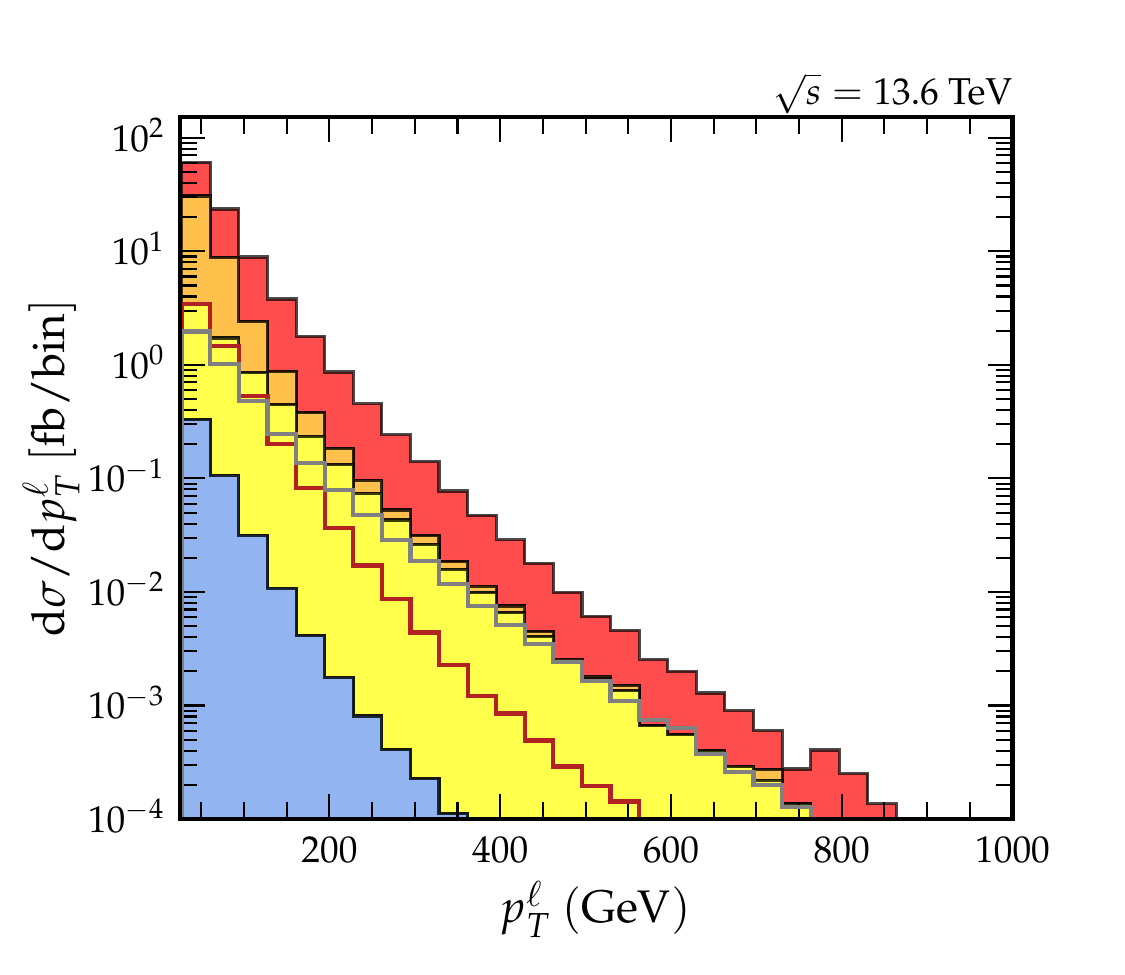}
\includegraphics[width=0.32\linewidth]{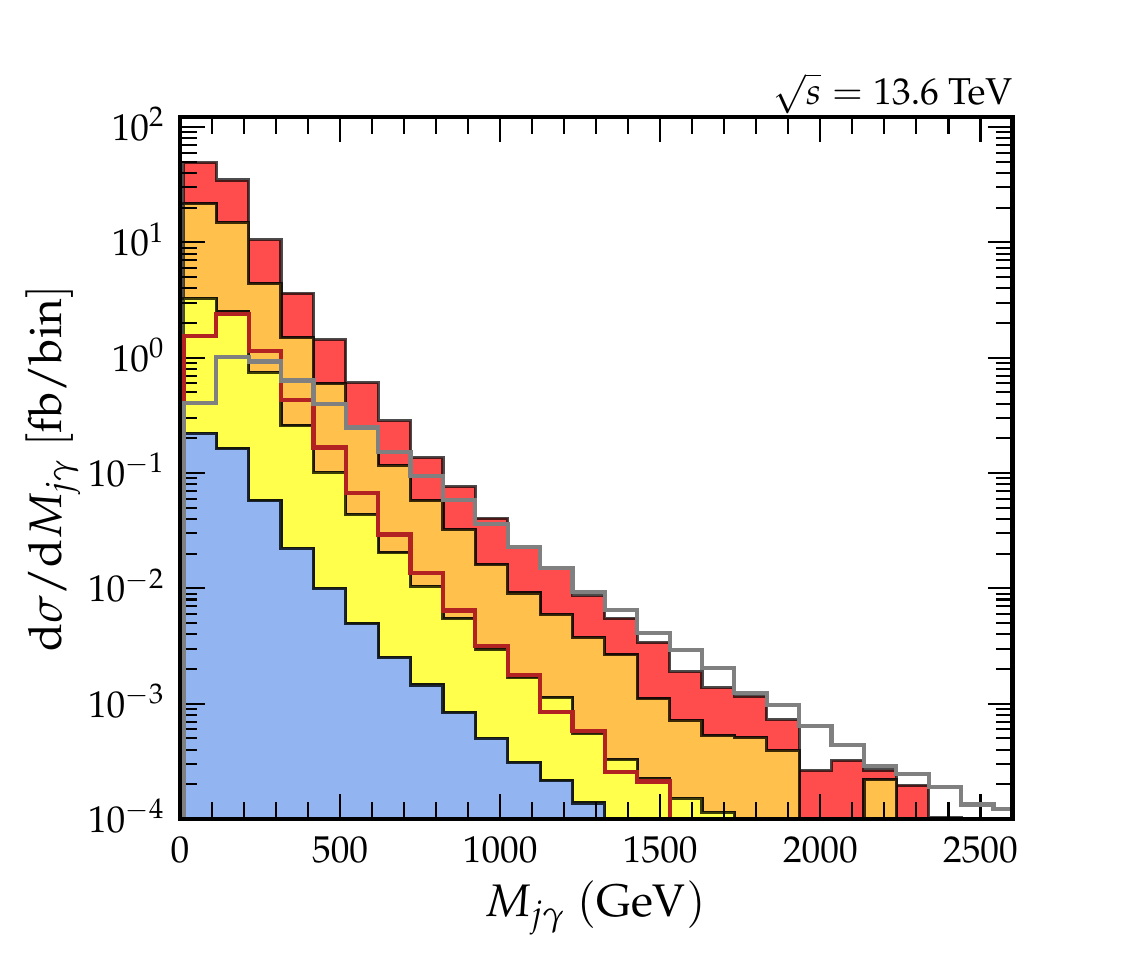}
\includegraphics[width=0.32\linewidth]{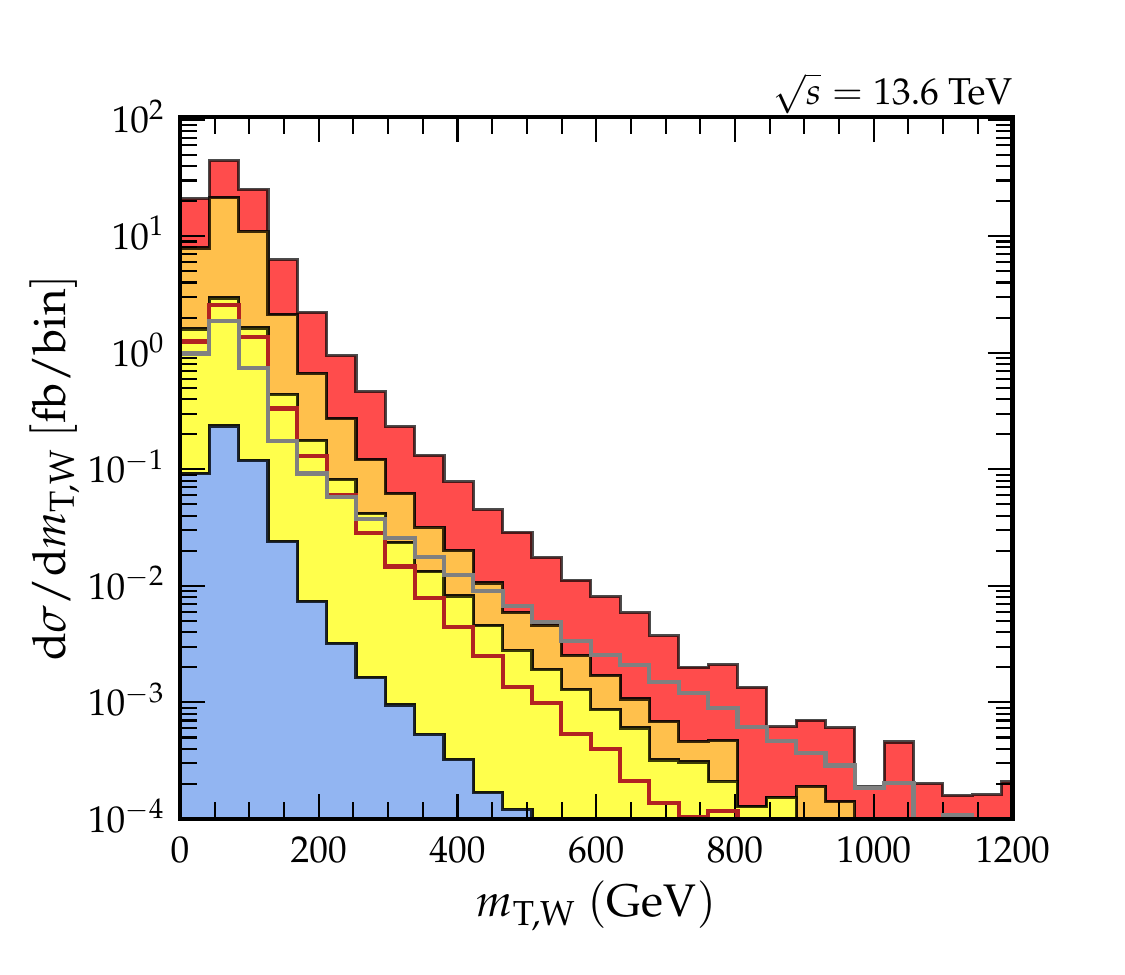}
\includegraphics[width=0.32\linewidth]{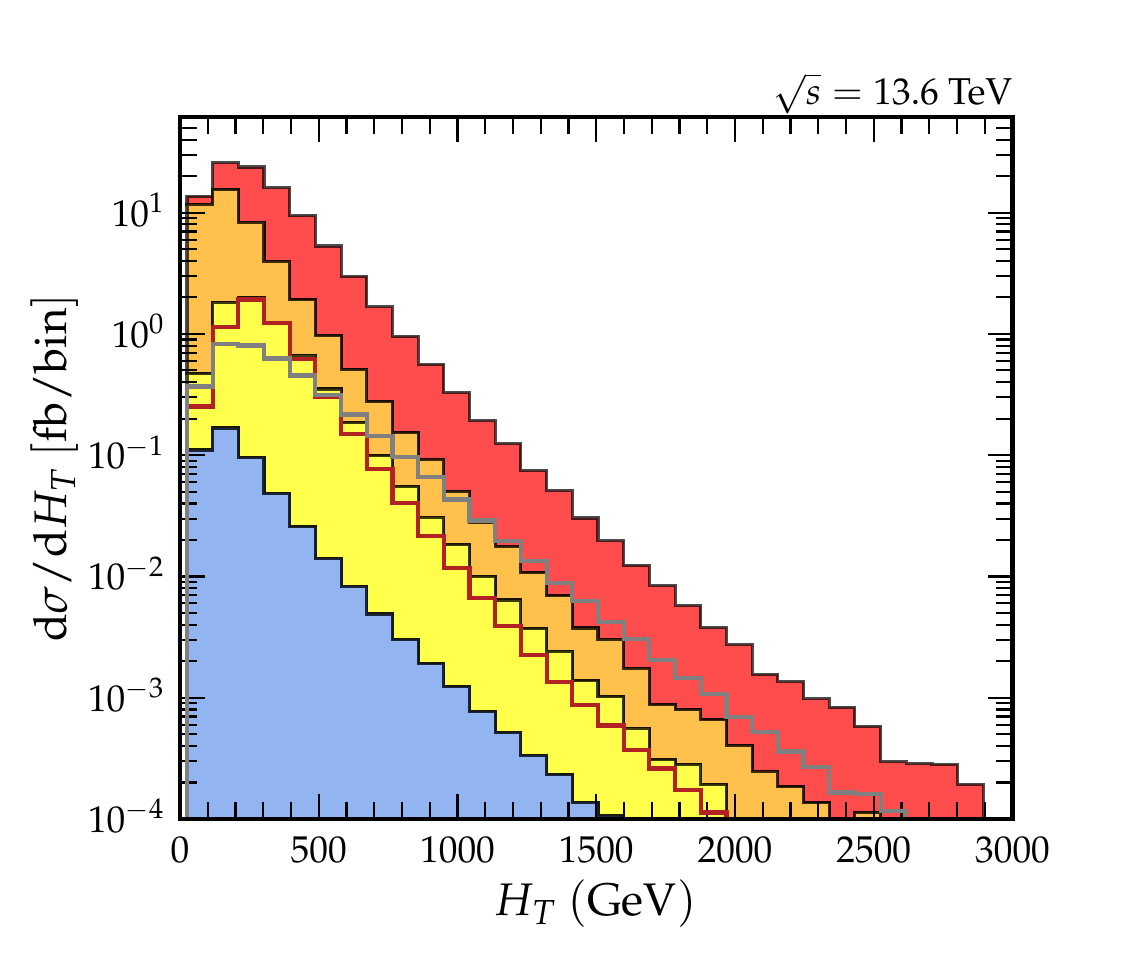}
\includegraphics[width=0.32\linewidth]{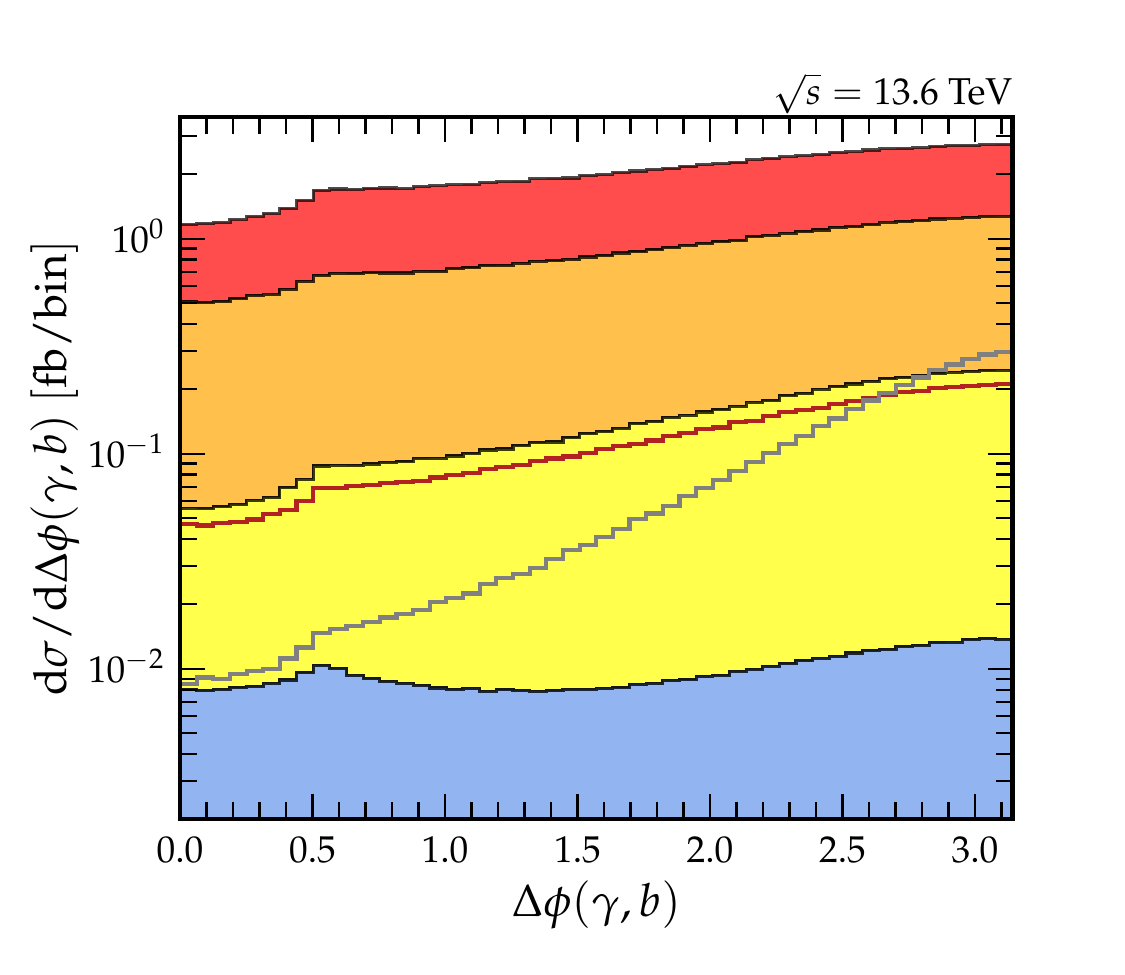}
\includegraphics[width=0.32\linewidth]{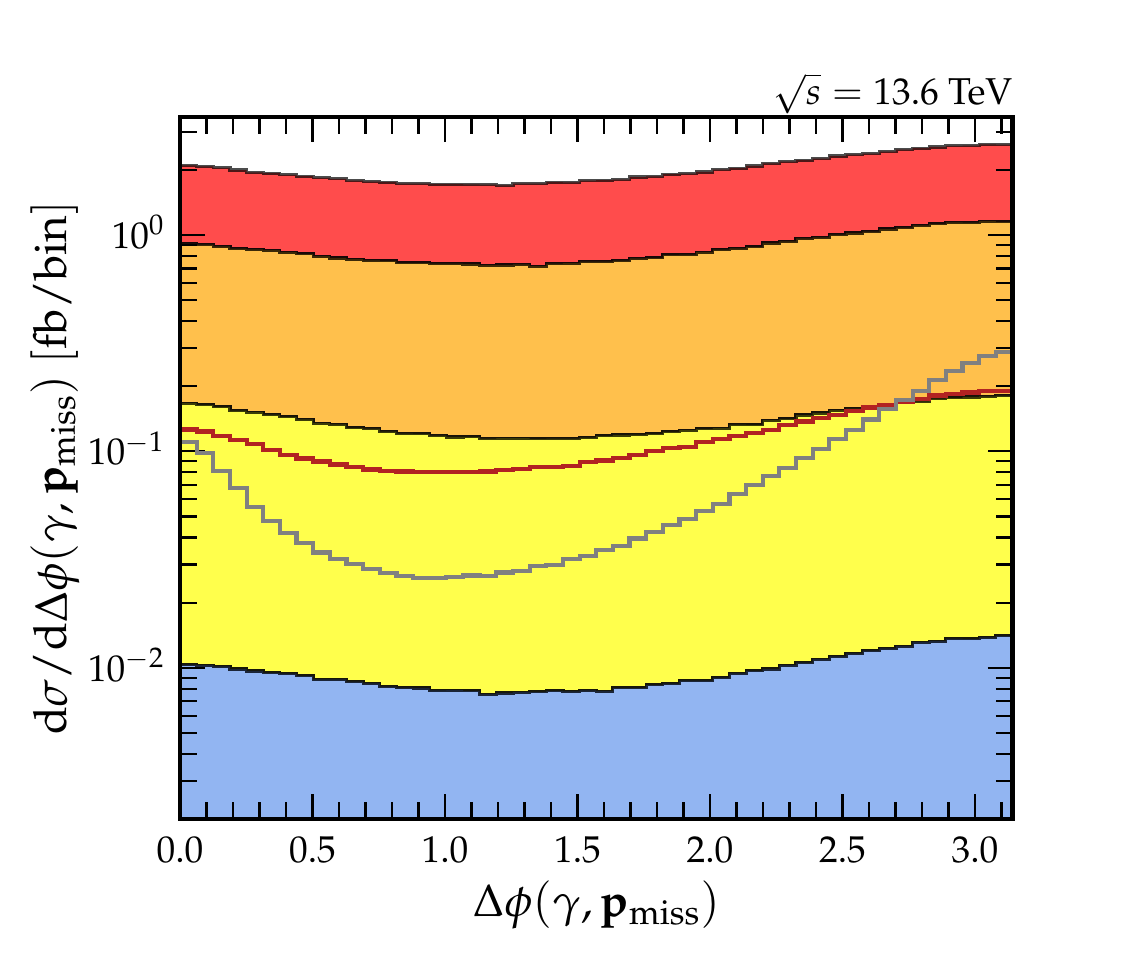}
\includegraphics[width=0.32\linewidth]{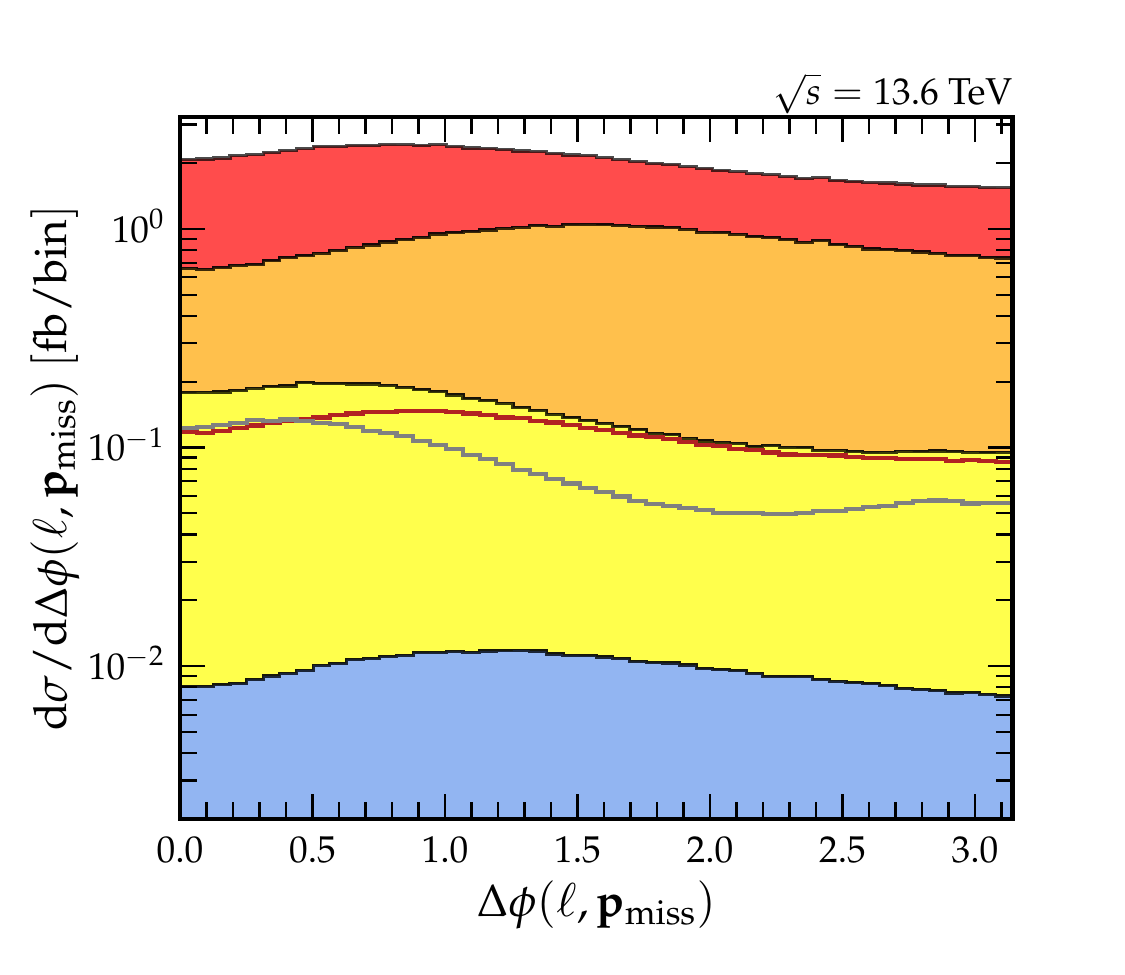}
\caption{Differential cross sections for representative kinematic variables used in the cut-based and MLP analyses of this study. Background contributions are stacked in ascending order of total cross section, while signal processes are overlaid as solid lines: single top production (grey) and top-pair production (red) for the benchmark coupling values $f_{tq}^\gamma = h_{tq}^\gamma = 3 \times 10^{-2}$. The top row displays the missing transverse energy (left), the transverse momentum of the signal photon (centre) and the transverse momentum of the signal lepton (right). The middle row shows the invariant mass of the jet-photon system (left), the transverse mass of the lepton-missing transverse momentum system (centre) and the scalar sum of jet transverse momenta (right). The bottom row presents azimuthal angle separations between the photon and the leading $b$-tagged jet (left), the photon and the missing transverse momentum (centre) and the lepton and the missing transverse momentum (right).}
\label{fig:histograms}
\end{figure}

%%%%%%%%%%%%%%%%%%%%%%%%%%%%%%%%%%%%%%%%
\subsection{Cut-based strategy}
\label{sec:cut:strategy}
%%%%%%%%%%%%%%%%%%%%%%%%%%%%%%%%%%%%%%%%
We initiate our cut-based analysis by applying a series of baseline preselection criteria to match the signal topology. Events are required to contain exactly one isolated charged lepton (either an electron or a muon) with transverse momentum $p_T > 27$~GeV and pseudo-rapidity $|\eta| < 2.5$, although for electrons we exclude the calorimeter transition regions of a typical LHC detector corresponding to $1.37 < |\eta_e| < 1.52$. Lepton isolation is enforced through the variable $I_\ell$, that we define as the scalar sum of the transverse momenta of all tracks $i$ separated by a transverse distance $\Delta R(i, \ell) < 0.3$ from the lepton,
\begin{equation}
    I_\ell = \sum_{i \in {\rm tracks}} p_T^i\,.
\end{equation}
To ensure isolation, electrons must satisfy $I_\ell < 5$~GeV while muons are required to meet $I_\ell / p_T^\ell < 0.1$. Photon isolation is treated analogously, using an isolation variable $I_\gamma$ defined in terms of the scalar sum of the transverse momenta of all tracks around the photon. Photons are then considered isolated if the following two conditions are satisfied,
\begin{equation}
    I_\gamma < \begin{cases}
        0.05 \times p_T^\gamma \, \qquad \qquad \quad {\rm for} \qquad \Delta R(i, \gamma) < 0.2\,, \\
        0.22 \times p_T^\gamma + 2.45 \, \qquad {\rm for} \qquad \Delta R(i, \gamma) < 0.4\,,
    \end{cases}
\end{equation}
and in our preselection procedure we require events to comprise exactly one isolated photon satisfying $p_T^\gamma > 20$~GeV and $|\eta^\gamma| < 2.37$ after excluding photons falling in the calorimeter transition region.

Jets are required to have $p_T > 25$~GeV and $|\eta| < 2.5$, this definition thereby rejecting forward jets, and events must contain at least one of such jet candidates. Among these jet candidates, we impose that at least one of them is $b$-tagged, which allows us to specifically target jets originating from the leptonic decay of top quarks. We remind that the detector response is parametrised following~\cite{Frank:2023epx}, thus using a $b$-tagging efficiency ($\mathcal{E}_{b \to b}$) and mistagging rates of a charm jet as a $b$-jet ($\mathcal{E}_{c \to b}$) and of a light jet as a $b$-jet ($\mathcal{E}_{j \to b}$) as a $b$-jet of
\begin{equation}\begin{split}
    \mathcal{E}_{j \to b} =&\ 10^{-2} + 3.8 \times 10^{-5} \times p_T\,, \\
    \mathcal{E}_{c \to b} =&\ 0.25 \times \tanh(0.018 \times p_T) \times (1 + 0.0013 \times p_T)^{-1}\,, \\
    \mathcal{E}_{b \to b} =&\ 0.85 \times \tanh(0.0025 \times p_T) \times \frac{25}{1 + 0.063 \times p_T}\,.
\end{split}\end{equation}
These functions yield an average $b$-tagging efficiency of approximately 70\% in the signal-dominated $p_T$ range, with typical mistagging rates of about 20\% for charm jets and 1\% for light-flavour jets. As will be shown, this has a notable impact on the sensitivity to top-photon FCNC interactions, leading to better performance for the charm-quark channel compared to the up-quark one.

Finally, the missing transverse momentum is computed as the negative vector sum of the transverse momenta of all visible objects in the event,
\begin{equation}
    {\bf p}_T^{\rm miss} = - \sum_i {\bf p}_{T, i}\,.
\end{equation}
In order to match the signal topology, we require the magnitude of this vector, $E_T^{\text{miss}} = |{\bf p}_T^{\rm miss}|$, to be greater than 30~GeV. 

\begin{table}
\setlength\tabcolsep{5pt}\renewcommand{\arraystretch}{1.2}
  \begin{center}
    \begin{tabular}{l cc cc cc}
    \toprule
 Process &   \multicolumn{2}{c}{Signal} & \multicolumn{2}{c}{$t\bar{t}\gamma$} & \multicolumn{2}{c}{$t\gamma$} \\ 
      & Events & $\epsilon$ [\%] & Events & $\epsilon$ [\%] & Events & $\epsilon$ [\%] \\ 
      \toprule
Initial & $6900.0$ & $-$ & $154341$ & $-$ & $142623$ & $-$ \\
$N_\ell = 1$ & 4151.19 $\pm$ 0.93 & 60.16 & 92055.78 $\pm$ 17.63 & 59.64 & 82098.46 $\pm$ 10.85 & 57.56 \\
$N_\gamma = 1$ & 3277.41 $\pm$ 0.84 & 47.50 & 49343.61 $\pm$ 12.27 & 31.97 & 40327.89 $\pm$ 6.91 & 28.28 \\
$N_{\rm jets} \geq 1$ & 3226.49 $\pm$ 0.83 & 46.76 & 49271.67 $\pm$ 12.26 & 31.92 & 39443.38 $\pm$ 6.75 & 27.66 \\
$N_{b} \geq 1$ & 1622.61 $\pm$ 0.50 & 23.52 & 36643.03 $\pm$ 9.65 & 23.74 & 22530.48 $\pm$ 3.97 & 15.80 \\
$E_{\rm T}^{\rm miss} > 30~{\rm GeV}$ & 1371.06 $\pm$ 0.43 & 19.87 & 31302.32 $\pm$ 8.43 & 20.28 & 18137.05 $\pm$ 3.23 & 12.72 \\
$\Delta R(\ell, j) > 0.4$ & 1333.99 $\pm$ 0.42 & 19.33 & 30945.66 $\pm$ 8.34 & 20.05 & 17957.06 $\pm$ 3.20 & 12.59 \\
$\Delta R(\gamma, j) > 0.5$ & 1273.19 $\pm$ 0.41 & 18.45 & 29549.59 $\pm$ 8.01 & 19.15 & 17223.96 $\pm$ 3.09 & 12.08 \\
$\Delta R(\gamma, \ell) > 0.5$ & 1265.24 $\pm$ 0.40 & 18.34 & 28796.77 $\pm$ 7.83 & 18.66 & 16799.76 $\pm$ 3.02 & 11.78 \\
Radiative $Z$ veto & 1241.34 $\pm$ 0.40 & 17.99 & 26854.47 $\pm$ 7.36 & 17.40 & 15583.64 $\pm$ 2.81 & 10.93 \\
$p_T^\gamma > 100~{\rm GeV}$ & 1031.37 $\pm$ 0.34 & 14.95 & 5494.69 $\pm$ 1.63 & 3.56 & 1859.43 $\pm$ 0.32 & 1.30 \\
      \toprule
    \end{tabular}
    \caption{Cutflow table for the signal process in the single production mode and the two dominant irreducible backgrounds $t\bar{t}\gamma$ and $t\gamma$. For each process, we report the number of events $n_i$ passing each selection criterion, along with the corresponding cumulative selection efficiency $\epsilon_i$ defined as $\epsilon_i = N_i / N_{\rm TOT}$ where $N_{\rm TOT}$ is the total (\textit{i.e.}\ initial) number of events for a given sample normalised to an integrated luminosity of $300$~fb$^{-1}$. The signal event yields are shown for a benchmark scenario with $f_{tq}^\gamma = h_{tq}^\gamma = 3 \times 10^{-2}$. \label{tab:efficiency}}
  \end{center}
\end{table}

Following the baseline preselection, we apply a small targeted set of signal-enhancing cuts designed to isolate events arising from top-photon FCNC interactions. To maintain a clean event topology and minimise overlap between reconstructed objects, we require a minimum angular separation between the photon, the lepton and all jets in the event. Specifically, we hence impose $\Delta R(\ell, j) > 0.4$, $\Delta R(j, \gamma) > 0.5$ and $\Delta R(\ell, \gamma) > 0.5$ for all signal jets. In the electron channel, we additionally veto events in which the invariant mass of the electron-photon system lies within the $Z$-boson mass window,
\begin{equation}
    81.2 < M_{e\gamma}/{\rm GeV} < 101.2\,,
\end{equation}
thus suppressing backgrounds from radiative $Z$-boson decays. At this stage of the analysis, we examine the key kinematic distributions shown in figure~\ref{fig:histograms}, which guide the final refinement of our signal region definition. Based on the photon transverse momentum spectrum, we hence impose an additional requirement on the signal photon of $p_T^\gamma > 100$ GeV. 

The cumulative impact of the above cut-based strategy is summarised in table~\ref{tab:efficiency}, which presents the cutflow for the signal process (only including the contributions of the single production mode, for illustrative purpose) alongside the dominant irreducible backgrounds, namely $t\bar{t}\gamma$ and $t\gamma$. After all selection requirements are applied, the signal efficiency reaches approximately $14.9\%$, while the corresponding efficiencies for the $t\bar{t}\gamma$ and $t\gamma$ backgrounds are found to be $3.56\%$ and $1.30\%$, respectively. The impact of other background contributions, such as those from $W\gamma$ + jets, are found to be much smaller, with efficiencies of the order of $10^{-5}$ or less. The resulting statistical significance $\mathcal{S}$ of the signal is then estimated using
\begin{equation}
    \mathcal{S} = \frac{N_s}{\sqrt{N_s + N_b + (\epsilon_b N_b)^2}} \approx 20.3~(2.66),
    \label{eq:significance}
\end{equation}
where $N_s$ and $N_b$ denote the \textit{total} number of signal and background events, respectively and $\epsilon_b$ is the background uncertainty. The above result is shown for two simple scenarios regarding background uncertainty; first we assume an ideal case where $\epsilon=0\%$ while the result inside bracket is shown for $\epsilon_b = 10\%$. Despite this relatively high significance, the corresponding signal purity in the selected region remains low, at around $10\%$, indicating that the signal region selects an event sample that is still dominated by background. This motivates the development of a more refined analysis strategy capable of enhancing signal purity while retaining good selection efficiency. However, such an optimisation is difficult to achieve with traditional cut-based methods, which are inherently limited in their ability to exploit the correlations between the considered observables and in the implementation of non-linear cuts. In the following sections, we explore how this challenge can be addressed using deep learning techniques, including multi-layer perceptrons, graph attention networks and transformer-based architectures.

%%%%%%%%%%%%%%%%%%%%%%%%%%%%%%%%%%%%%%
\subsection{Deep-learning strategies}\label{sec:DL}
%%%%%%%%%%%%%%%%%%%%%%%%%%%%%%%%%%%%%%%
\subsubsection{Generalities}
In this section, we improve our analysis of top-photon FCNC interactions using three deep learning techniques: multi-layer perceptrons (MLPs), graph attention networks (GATs) and transformer-based networks\footnote{In this work, signal and background samples simulated at leading order in QCD are used for both the training and validation stages. We stress out that higher-order predictions may induce shifts in the network latent space distribution due to impact on the kinematic distributions and their correlations. To mitigate such effects, the network architecture can be adapted using domain-adversarial techniques, which improve the generalisation of the latent space and enhance robustness against bias. This strategy has been discussed, for example, in \cite{Baalouch:2019fhm,Bishara:2023epi}. Whereas advanced architectures with a large number of tunable parameters, such as graph neural networks and transformers, are particularly well suited to incorporate this and address the issue effectively, such an explorative study is left for future work.}. These networks have been chosen for their complementary strengths in handling the input data, which is structured as a set of features derived from the reconstructed kinematic properties of the final-state objects in each event. While MLPs operate on fixed-size vectors and treat all features independently, GATs and transformers can process richer and more complex event representations that preserve the relationships between the objects within an event.

MLPs are networks composed of fully connected layers with non-linear activation functions that allow for the modelling of intricate dependencies among the input variables~\cite{Rumelhart:1986, Goodfellow:2016}. They are especially effective at learning smooth and continuous decision boundaries in high-dimensional feature spaces, making them a natural baseline for collider event classification due to their capacity for generalisation. In our setup, each event is represented as a fixed-size feature vector with each component encoding a high-level or low-level observable. For a dataset of $N$ events, the input to our ML-based analysis toolchain is a two-dimensional array of shape $(\text{batch size}, \text{number of features})$, where the batch size represents the number of events processed simultaneously during training or inference, and the number of features corresponds to the number of kinematic variables considered. Each row in this array corresponds to one event given as a flattened feature vector and each column refers to a specific observable. These features are however treated as independent inputs; thus, MLPs do not exploit any intrinsic spatial or relational structure in the event and instead infer correlations between features purely through learned transformations across its dense layers.

GATs, by contrast, are designed to incorporate the relational structure between the reconstructed objects within an event~\cite{Velickovic:2017lzs, Yu:2023juh, Calafiura:2024qhv, Esmail:2023axd}. In this approach, each event is encoded as a graph where nodes correspond to final-state objects and edges reflect their pairwise relationships. This structure allows the model to capture spatial correlations and interaction patterns that cannot be explicitly modelled by MLPs. As a result, GATs provide a more expressive and flexible framework for learning patterns that depend not only on individual object properties but also on how these objects relate to one another within the event. The input is formalised as a batch of graphs $G = (V, E)$ where $V$ is the set of nodes (\textit{i.e.}\ objects) each associated with a feature vector, and $E$ is the set of edges encoding inter-object relationships. These graphs are thus represented as two tensors: one of shape $(\text{batch size}, \text{number of nodes}, \text{node feature dimension})$ and another of shape $(\text{batch size}, \text{number of edges}, \text{edge feature dimension})$, thereby allowing efficient batched processing. In our analysis, we construct graphs with nine nodes per event corresponding to the five leading jets, the leading photon, the leading lepton, the reconstructed leptonically-decaying top quark and the missing transverse energy. Each node carries a ten-dimensional feature vector encoding physical quantities such as its transverse momentum $p_T$, pseudo-rapidity $\eta$, azimuthal angle $\phi$, energy $E$, electric charge $Q$ and a one-hot encoded object identifier. However, unlike MLPs, GATs can dynamically adjust to varying object multiplicities and event topologies.

Transformer networks, originally developed for natural language processing tasks using an encoder-decoder architecture~\cite{Vaswani:2017lxt}, have recently been adapted for collider event classification~\cite{Qu:2022mxj, Wu:2024thh, Hammad:2023sbd, Hammad:2024cae, Hammad:2024hhm}. In our context, we adopt an encoder-only architecture inspired by the particle cloud formalism~\cite{Komiske:2018cqr, Qu:2019gqs}, where each event is treated as an unordered set of objects akin to a point cloud in computer vision. This permutation-invariant representation allows transformer encoders to dynamically assess the relevance of each final-state object with respect to all others in the event using a self-attention mechanism, effectively capturing both short-range and long-range interrelations and providing a powerful framework for modelling complex object dynamics within collider events. In practice, each object is considered as a token, and the input is formatted as a tensor of shape $(\text{batch size}, \text{number of objects}, \text{number of features})$. As with our GAT-based analysis, object features include kinematic variables, electric charge and a one-hot identifier for particle type. Moreover, the number of objects per event is capped at nine, with padding or truncation applied as necessary to ensure uniformity for batch processing. The self-attention layers then dynamically compute contextualised representations of each object by aggregating information from all others, enabling the model to extract both global and local patterns relevant for signal-background discrimination.

In the next subsections, we detail the architecture, training procedure and performance of each considered deep-learning model. A comparative evaluation of their classification power is then presented, highlighting the advantages of each approach in capturing top-photon FCNC effects.

%%%%%%%%%%%%%%%%%%%%%%%%%%%%%%%%%%%%%%%
\subsubsection{Methodology}
%%%%%%%%%%%%%%%%%%%%%%%%%%%%%%%%%%%%%%
Training and evaluating deep neural networks for classification tasks in collider physics involves a systematic pipeline integrating statistical learning, numerical optimisation and computational infrastructure. Once the datasets are constructed, each event is assigned a binary label: signal events are given $Y=1$ and background events $Y=0$. These samples are then merged into a single dataset, and a random shuffle is applied to remove any ordering-related biases before training.

The learning process is iterative. In each training epoch defined as one complete pass over the dataset, the model updates its internal parameters to minimise a predefined loss function. Input events, represented as vectors, graphs or sets depending on the architecture, are fed into the network in mini-batches during the forward pass. Each input propagates through the network layers, and the output layer produces class scores that are converted into probabilities using a softmax activation function. To quantify the agreement between predicted probabilities and the true labels, we use a binary cross-entropy loss function where lower loss values indicate better predictive accuracy on the training data. After computing the loss, gradients are calculated via backpropagation, which applies the chain rule of calculus to evaluate how each weight and bias contributes to the final error. These gradients guide the update of the model parameters using stochastic gradient descent or an adaptive optimiser such as AdamW~\cite{loshchilov2017decoupled}. This optimisation cycle comprising forward propagation, loss evaluation, backpropagation and parameter update is repeated over multiple epochs until convergence. The trained model is then evaluated on an independent test set to assess its generalisation performance on unseen data.

\begin{table}\renewcommand{\arraystretch}{1.2}
    \setlength\tabcolsep{0.35 cm}
    \begin{tabular}{lccc}
          & MLP  & GAT & Transformer\\
        \toprule
        Input layer dimension & (29)  & (9,10) & (9,10)   \\
        Number of hidden layers & 5 FC & 10 GAT & 10 self-attention \\
        Feature embedding layers & $-$ & [512,256,128] FC& [512,256,128]FC   \\
        Output layer dimension & 1 neuron & 1 neuron & 1 neuron \\
        Output layer activation & Sigmoid & Sigmoid & Sigmoid \\
        Number of attention heads& $-$ & 32 & 16 \\
        Drop out rate & $5\%$ & $25\%$ & $25\%$ \\
        Pooling & $-$ & Average & Average \\
        Final MLP & $-$ & [128,64] FC &[128,64] FC\\
        Trainable parameters & 189K &1.20M &2.01M\\
        Loss function & BCE& BCE & BCE \\
        Optimiser & AdamW  & AdamW & AdamW \\
        Initial learning rate & 0.001 & 0.0005& 0.0005 \\
        Learning rate scheduler & Cosine Annealing& Cosine Annealing&Cosine Annealing\\ 
        Early stopping patience & 5 epochs & 5 epochs &5 epochs \\
        Epochs & 20 & 23  &21 \\
        Batch size & 128 & 128&128  \\
        \hdashline 
        Deep learining framework &\multicolumn{3}{c}{PyTorch Lightning}\\ 
        Training device &\multicolumn{3}{c}{GPU}\\
        Computational cost  &\multicolumn{3}{c}{2 $\times$ NVIDIA Quadro RTX 6000}\\
        \bottomrule
    \end{tabular}
    \caption{\label{tab:structure}%
        Hyperparameters of the different considered deep learning networks. The abbreviation `FC' refers to fully connected layers while the abbreviation `BCE' denotes binary cross-entropy loss.}
\end{table}

Optimisation is a crucial step in the development of deep neural networks and is aimed at identifying the network configuration that yields optimal performance. This process involves tuning hyperparameters, namely parameters whose values are not learned during training but must be set prior to the optimisation process. Examples include the learning rate, batch size, number of layers or dropout rate. These hyperparameters have a significant impact on the model convergence, generalisation and training stability, which makes their selection a critical component of model design. A standard approach to hyperparameter tuning is the grid search method, which exhaustively evaluates the model performance over all possible combinations of predefined hyperparameter values. While systematic, grid search is computationally expensive and scales poorly with the number of hyperparameters and value ranges considered. To mitigate this issue, we employ a more efficient alternative known as a random grid search. In this approach, hyperparameter values are sampled from predefined statistical distributions rather than fixed grids, allowing broader exploration of the hyperparameter space with significantly reduced computational cost.

For the considered MLP architecture, we varied the number of neurons per layer in the range $\{64, 128, 256, 512\}$ and the dropout rate from 5\% to 20\%. For the GAT and transformer models, the hyperparameter scan additionally included the number of attention layers, number of attention heads, learning rate and batch size. A summary of the optimal hyperparameter configurations used in our analysis is provided in table~\ref{tab:structure}.

Finally, we remind that another source of variability in performance arises from fluctuations induced by the random partitioning of data into training, validation and test sets. To reduce this statistical variance and obtain a robust estimate of the model's generalisation ability, we employ $k$-fold cross-validation. In this method, the dataset is divided into $k$ equal subsets. The model is then trained on $k{-}1$ of them and validated on the remaining one. This procedure is repeated $k$ times, with each subset serving once as the validation set. The final performance metric is then obtained by averaging over all $k$ runs. For all models in our study, we use $k=3$, ensuring consistent and statistically stable evaluation across architectures.

%%%%%%%%%%%%%%%%%%%%%%%%%%%%%%%%%%%%%%%
\subsubsection{Multi-Layer Perceptron}
%%%%%%%%%%%%%%%%%%%%%%%%%%%%%%%%%%%%%%
Starting from a combination of low- and high-level kinematical observables, we employ an MLP to develop a classification strategy that outperforms the cut-based approach of section~\ref{sec:cut:strategy} by capturing characteristic patterns in both signal and background event samples. The fully connected structure of the MLP makes it then well suited for learning global correlations and translating them into a classification decision. In particular, the MLP is capable of identifying non-linear patterns that increase the separation power between signal and background~\cite{Rumelhart:1986, Goodfellow:2016}.

A total of 29 input features are used. These include the four-momentum components of the final-state charged lepton ($p_T^\ell$, $\eta_\ell$, $\phi_\ell$, $E_\ell$), photon ($p_T^\gamma$, $\eta_\gamma$, $\phi_\gamma$, $E_\gamma$) and leading $b$-tagged jet ($p_T^b$, $\eta_b$, $\phi_b$, $E_b$), as well as the missing transverse energy ($\met$) and its azimuthal angle $\phi_{\rm miss}$. We also include the number of small-$R$ jets (both $b$-tagged and light), several azimuthal separations between final-state objects ($\Delta\phi(\gamma, \ptmiss)$, $\Delta\phi(\ell, \ptmiss)$, $\Delta\phi(b, \ptmiss)$, $\Delta\phi(\gamma, b)$, $\Delta\phi(\gamma, \ell)$, $\Delta\phi(\ell, b)$) and their angular distances ($\Delta R(\gamma, b)$, $\Delta R(\gamma, \ell)$, $\Delta R(b, \ell)$), where $\Delta\phi(i, j)$ is normalised between 0 and $\pi$. In addition to these low-level observables, a set of high-level features is introduced to aid the classification. These include the transverse mass of the $W$ boson system made of the lepton and the missing momentum,
\begin{equation}
    m_{\rm T,W} = 2  p_T^\ell  \met \sqrt{1 - \cos(\Delta\phi(\ell, \ptmiss)}\,,
\end{equation}
the scalar sum of the transverse momenta of all jets in the event,
\begin{equation}
    H_T = \sum_{i = 1}^{N_{\rm jets}} p_T^i\,,
\end{equation}
and the event’s effective mass,
\begin{equation}
    m_{\rm eff} = H_T + \met + p_T^\ell + p_T^\gamma\,.
\end{equation}
We finally also include the invariant mass of the system formed by the photon and the leading untagged jet, along with the sign of the charged lepton's electric charge $Q_\ell$.

\begin{figure}
    \centering
    \includegraphics[width=0.9\linewidth]{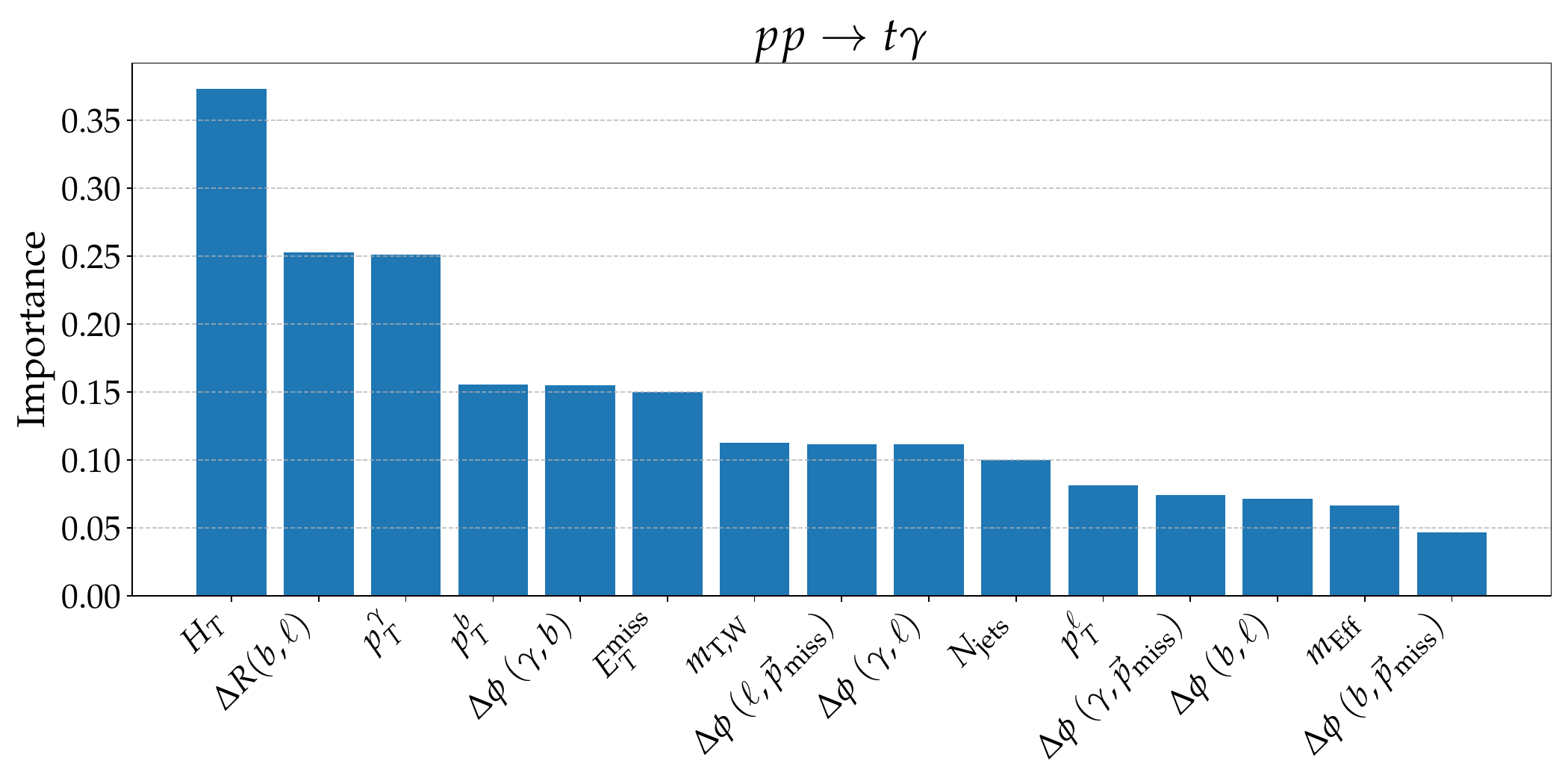}
    \caption{Feature importance for the top 15 ranked input variables used in the MLP network. The $y$-axis shows the increase in model loss when each variable is randomly shuffled, averaged over 10 repetitions.}
    \label{fig:feature:importance}
\end{figure}

The relative importance of the most influential features is illustrated in figure~\ref{fig:feature:importance}. We estimate the contribution of each feature using a permutation-based method in which a single variable is randomly shuffled, thus breaking its correlation with other inputs, while the others are left intact. The resulting increase in model loss is computed and averaged over ten trials, the final score thus reflecting the difference between the average shuffled loss and the baseline loss. This analysis reveals that $H_T$, $\Delta R(b, \ell)$, and $p_T^\gamma$ are the most impactful variables for distinguishing signal from background.

Although the MLP achieves high classification performance, certain kinematic variables exhibit similar distributions for both signal and background, thereby limiting separation power. To mitigate this, we apply selection cuts optimised to increase the signal-to-background ratio before training the MLP, that we choose as the cuts introduced in section~\ref{sec:cut:strategy}. In principle, the use of preselection could introduce performance degradation due to intrinsic inter-variable correlations: removing or cutting on one variable may influence others. However, we find that these correlations do not significantly impact performance: removing the most correlated inputs from the training set leads to only minor degradation in classification metrics.

The adopted MLP architecture consists of an input layer with 29 neurons corresponding to the selected features. This is followed by four fully connected layers with 512, 256, 128 and 64 neurons respectively, each using a GeLU activation function. Dropout layers with a rate of 10\% are placed after each hidden layer to mitigate overfitting, and the final layer is a single-neuron output with a sigmoid activation function providing a probability estimate for signal classification. For training, we use the AdamW optimiser~\cite{loshchilov2017decoupled} with a learning rate of $10^{-3}$ and a weight decay of $10^{-4}$ to minimise a binary cross entropy loss function.

\begin{figure}
    \centering
    \includegraphics[width=0.99\linewidth]{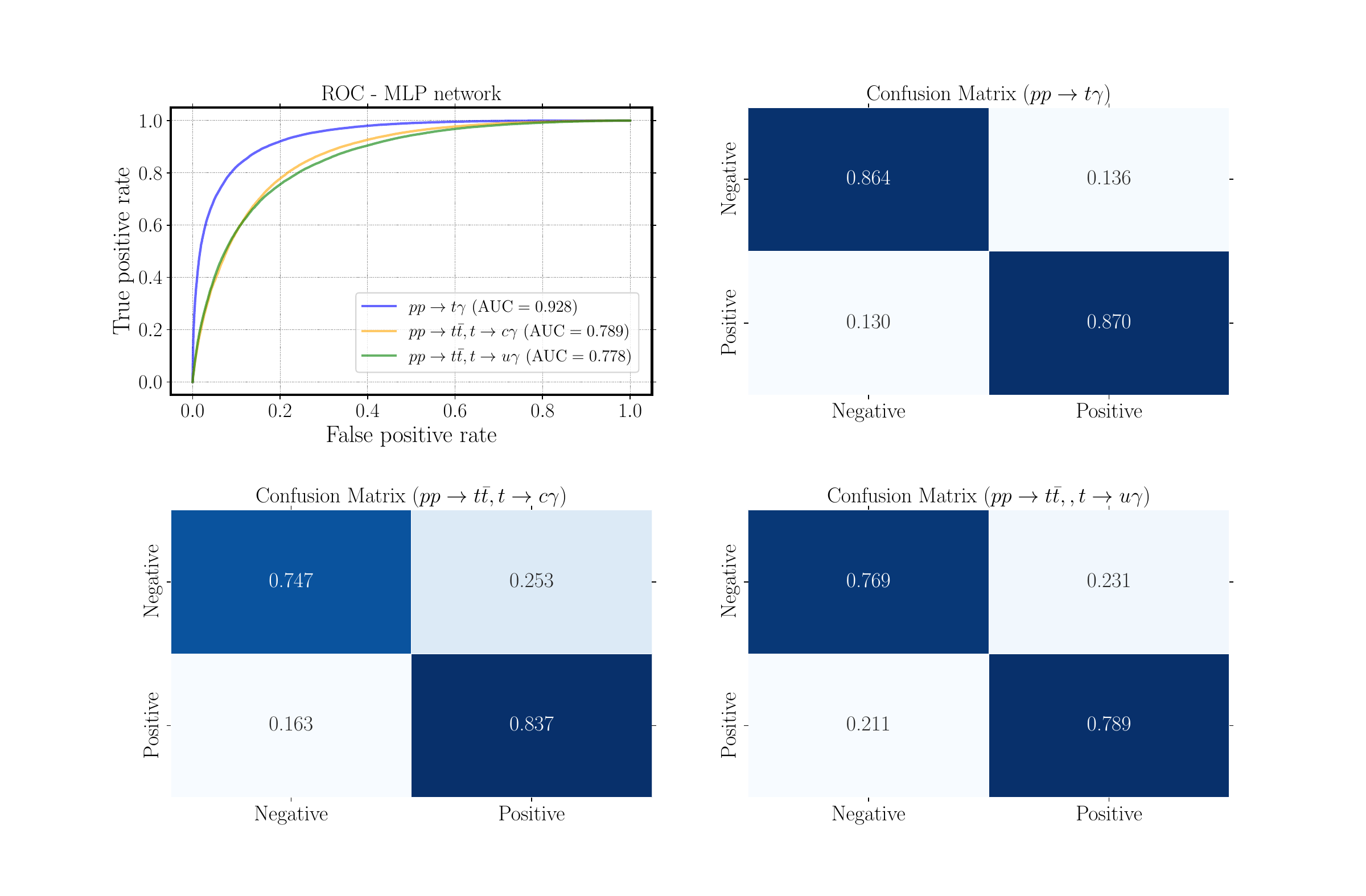}
    \caption{Classification performance of our MLP network. We present ROC curves for $pp\to t\gamma$ (blue), $pp\to\bar{t}t$ with $t\to u\gamma$ (green) and $pp\to\bar{t}t$ with $t\to c\gamma$ (orange) in the top-left panel. The remaining panels show the corresponding confusion matrices.}
    \label{fig:MLP_resutls}
\end{figure}

The classification performance of the MLP is summarised in figure~\ref{fig:MLP_resutls} in which we show associated confusion matrices and receiver operating characteristic (ROC) curves. They both consist in valuable tools for evaluating classification models, though they serve different purposes and provide complementary insights. The confusion matrix offers a detailed, threshold-specific breakdown of predictions by reporting the number of true positives, true negatives, false positives and false negatives. This facilitates a granular understanding of the model's performance in real-world decision-making scenarios, particularly in assessing the types of errors made. Derived metrics such as precision, recall and the F1-score are especially important when the misclassification costs are asymmetric~\cite{Cornell:2021gut}. By contrast, the ROC curve provides a threshold-independent evaluation by plotting the true positive rate against the false positive rate across all possible classification thresholds. It gives a global view of the model's discriminative ability. However, a high ROC score given through the area under the curve (AUC) does not guarantee optimal performance at any specific threshold, where the model might still yield a high rate of false positives or miss true signals. Therefore, while the ROC curve summarises the overall classification potential, the confusion matrix remains essential for fine-tuning the model's behaviour at specific operating points relevant to practical applications. In the top-left panel of figure~\ref{fig:MLP_resutls}, we show results for the three separate signal processes: $pp\to t\gamma$ and $pp\to\bar{t}t$ production followed by a rare $t\to u \gamma$ or $t\to c \gamma$ decay. The ROC curves indicate that the MLP achieves the highest discrimination for the single-top channel, with an AUC of 0.928. The $t\bar{t}$ production modes yield slightly lower but comparable performance, with AUCs of 0.789 and 0.778 for the charm and up-quark FCNC top decays respectively. The confusion matrices  shown in the other three panels of the figure support these observations and confirm the consistent behaviour of the classifier across the different signal channels.

%%%%%%%%%%%%%%%%%%%%%%%%%%%%%%%%%%%%%%
\subsubsection{Graph Attention Network}
%%%%%%%%%%%%%%%%%%%%%%%%%%%%%%%%%%%%%%
Graph Neural Networks (GNNs) are powerful deep learning architectures designed to process graph-structured data. In the context of high-energy physics, such graphs often represent the relationships between final-state particles, with nodes corresponding to individual objects and edges encoding pairwise interactions or geometrical proximity. In classification tasks, GNNs aim to learn representations of signal and background events by aggregating information from each node's neighbourhood. This aggregation captures both local graph structure and node features, which allows the model to distinguish between different classes of events. Among the various GNN architectures, Graph Attention Networks (GATs) have demonstrated particularly strong performance in particle physics applications~\cite{Velickovic:2017lzs, Yu:2023juh, Calafiura:2024qhv, Esmail:2023axd}. Traditional GNNs use fixed or uniform weighting schemes during neighbourhood aggregation. However, this uniformity can limit the model's ability to differentiate between relevant and irrelevant connections~\cite{Shlomi:2020gdn, Thais:2022iok, Esmail:2024gdc, Esmail:2024jdg, Hammad:2025ewr}. In contrast, GATs introduce an attention mechanism that dynamically assigns weights to neighbouring nodes during message passing. This enables the network to focus on the most informative parts of the graph for each node, which is particularly advantageous in heterogeneous or noisy environments where not all neighbours contribute equally. As a result, GATs offer improved interpretability and adaptability, often leading to higher classification accuracy and robustness in practical applications.

In GATs, the attention mechanism computes a pairwise attention score for each connected node pair, quantifying how much a given neighbour contributes to the target node's feature update~\cite{Esmail:2023axd}. Specifically, the attention coefficient between node $i$ and its neighbour $j$ is defined as
\begin{equation}
e_{ij} = \text{LeakyReLU} \left( a^\top \left[ W h_i \oplus W h_j \right] \right),
\end{equation}
where $h_i$ and $h_j$ denote the input feature vectors of nodes $i$ and $j$, respectively. The weight matrix $W \in \mathbb{R}^{F' \times F}$ transforms the input features to a higher-dimensional representation, and the $\oplus$ operation denotes vector concatenation. In addition, the vector $a \in \mathbb{R}^{2F'}$ is a trainable attention vector and the LeakyReLU activation function introduces non-linearity. Finally, we emphasise that this formulation treats both nodes symmetrically to ensure permutation invariance. The raw attention scores are then normalised using a softmax function over the neighbourhood $\mathcal{N}_i$ of node~$i$, 
\begin{equation}
\alpha_{ij} = \text{softmax}_j(e_{ij}) = \frac{\exp(e_{ij})}{\sum_{k \in \mathcal{N}_i} \exp(e_{ik})},
\end{equation}
where $\alpha_{ij}$ represents the normalised attention coefficient from node $j$ to node $i$. Once these coefficients are computed, the final node representation is obtained via a weighted sum of the transformed features of neighbouring nodes,
\begin{equation}
\tilde{h}_i = \sigma \left( \sum_{j \in \mathcal{N}_i} \alpha_{ij} W h_j \right),
\end{equation}
where $\sigma(\cdot)$ is an activation function, typically ReLU or ELU, applied to the aggregated feature vector. 

\begin{figure}
    \centering
    \includegraphics[width=.8\linewidth]{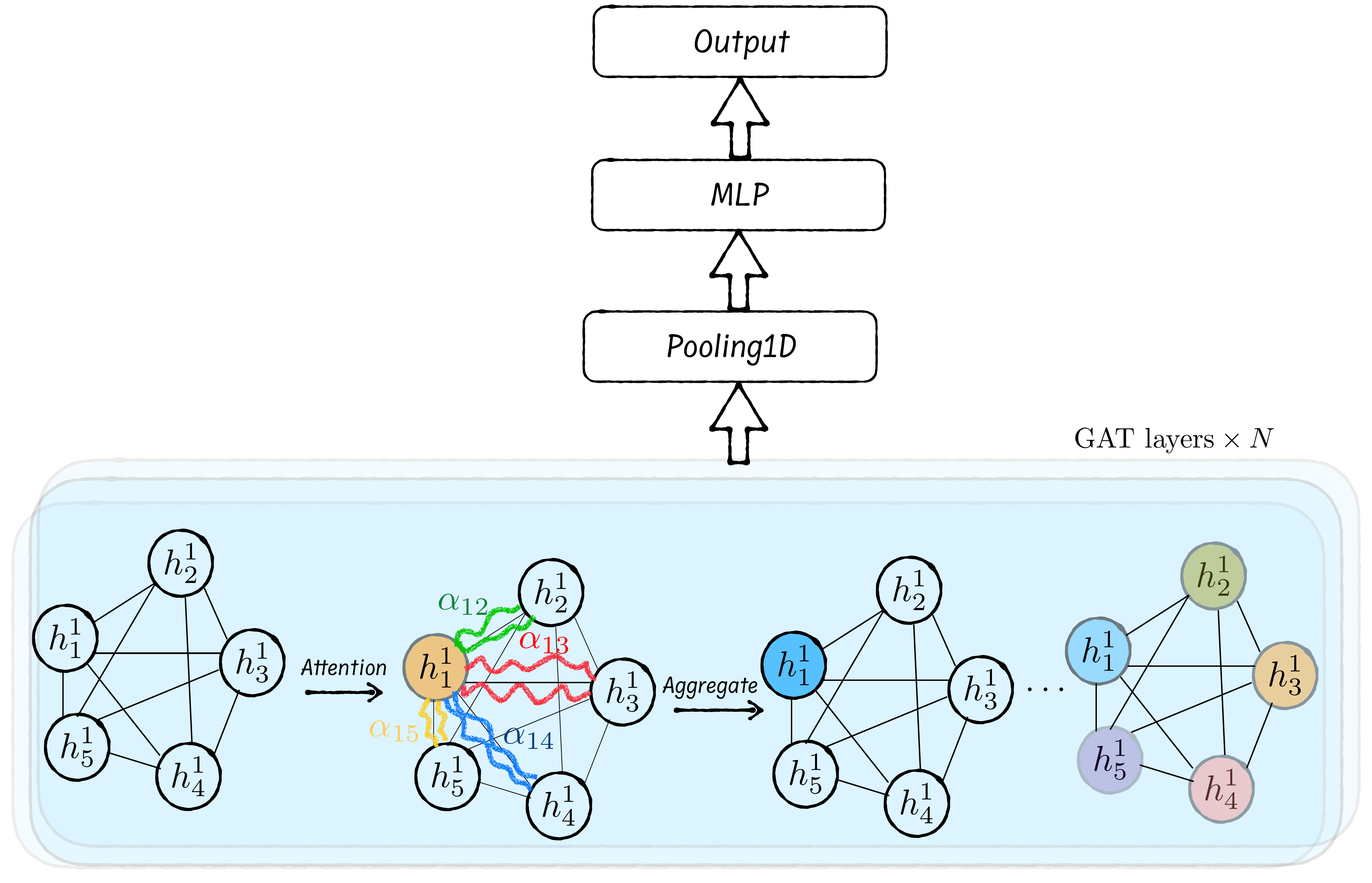}
    \caption{Schematic illustration of the employed GAT network architecture.\label{fig:GAT_network}} \vspace{.5cm}
    \includegraphics[width=0.95\linewidth]{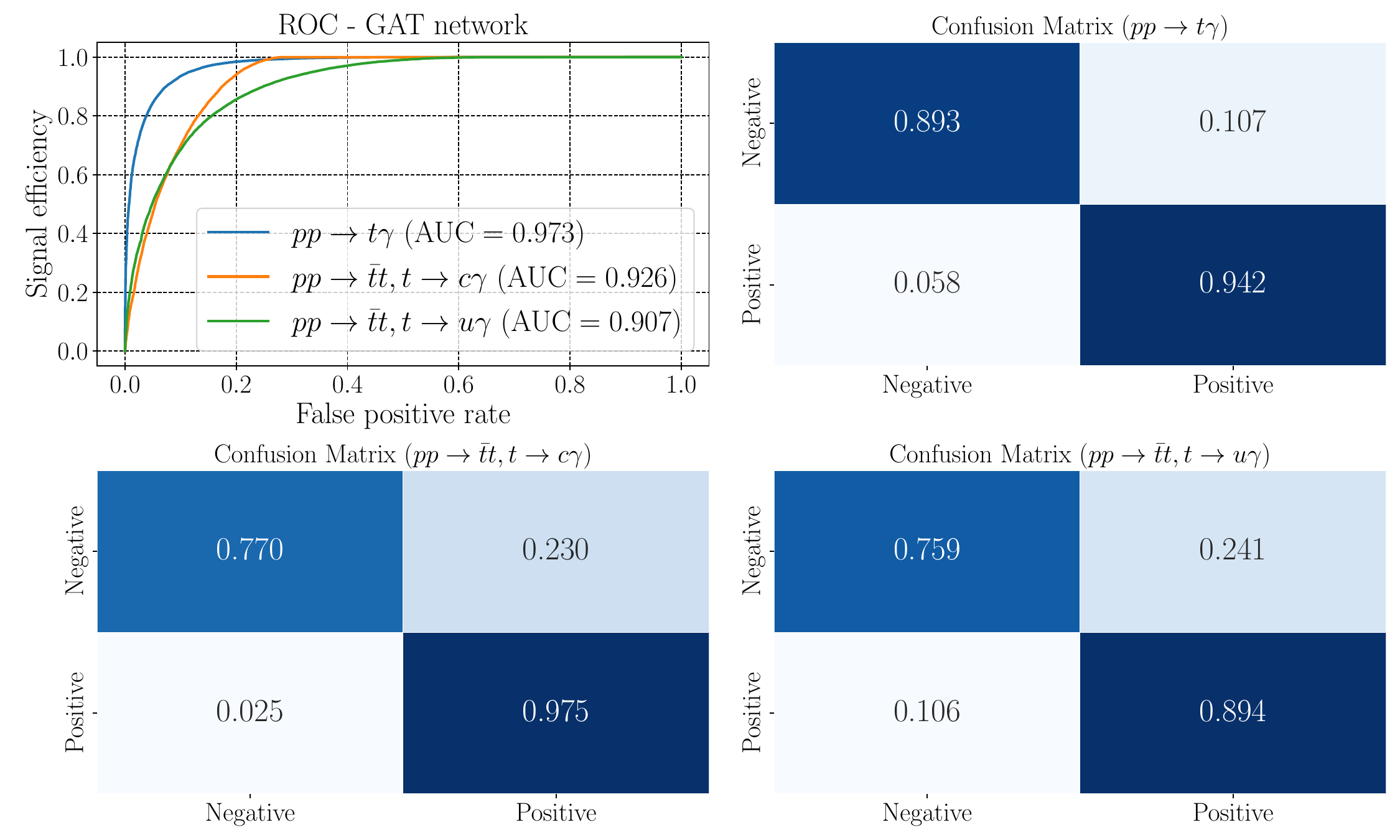}
    \caption{Classification performance of the GAT network shown as ROC curves (top-left) for $pp \to t\gamma$ (blue), $pp \to \bar{t}t$ with a $t \to u \gamma$ decay (orange) and a $t \to c \gamma$ decay (green), as well as through the corresponding confusion matrices (presented in the other panels).\label{fig:GAT_resutls}}
\end{figure}
The GAT network architecture employed in this work comprises 10 layers, each with 32 attention heads. The outputs from the attention heads are averaged and passed to a fully connected MLP consisting of two dense layers with 128 and 64 neurons respectively, and ReLU activation functions. The final output layer then contains two neurons with a softmax activation function, resulting in a total of approximately 1.2 million trainable parameters. A schematic overview of the entire GAT network used is shown in figure~\ref{fig:GAT_network}. Finally, training is performed for up to 23 epochs using a batch size of 256, with early stopping triggered after 5 consecutive epochs without improvement.

Figure~\ref{fig:GAT_resutls} summarises the classification performance of the GAT network across the three considered processes: $pp \to t\gamma$, $pp \to \bar{t}t$ with a $t \to u \gamma$ decay and a $t \to c \gamma$ decay. The ROC curves reveal that the single top production process ($pp \to t\gamma$) achieves, like for our MLP implementation, the highest discrimination power with an AUC of $0.973$. Furthermore, the top pair production processes yield slightly lower AUC scores of $0.907$ and $0.926$ respectively, which is this time much larger than in the MLP case. Finally, we note that the confusion matrices reflect a consistent behaviour, showing a reliable separation of signal and background across all processes.  

%%%%%%%%%%%%%%%%%%%%%%%%%%%%%%
\subsubsection{Transformers}%%
%%%%%%%%%%%%%%%%%%%%%%%%%%%%%%
Transformer models, first introduced in \cite{vaswani2017attention}, have revolutionised deep learning by enabling highly parallel processing and effectively capturing long-range dependencies. In particular, their application in high-energy physics have already shown promising results in analysing complex particle interactions~\cite{Qu:2022mxj, Wu:2024thh, Hammad:2023sbd, Hammad:2024cae}. Unlike other neural networks, transformers rely on a multi-head self-attention mechanism that enables global context modelling across input elements. This makes them particularly suitable for studying the intricate and highly correlated structure of particle clouds \cite{Qu:2019gqs} in collider events and assess the contribution of each particle to the overall event structure to allow for improved signal and background classification.

In this work, we adapt a transformer encoder to analyse the spatial and kinematic properties of individual particles within an event while preserving their contextual relationships by representing each final-state object as an input token. The input to the network is structured as an unordered set of objects,
\begin{equation}
X = x_1^i, x_2^i, \dots, x_n^i,
\label{eq:4.14}
\end{equation}
where $i$ denotes the number of features per object and $n = 9$ is the fixed number of input objects, corresponding to the five leading
jets, the leading photon, the leading lepton, the reconstructed leptonically-decaying top quark and the missing transverse energy. Each object's features are projected into three vectors, named Query ($Q$), Key ($K$) and Value ($V$), via learnable linear transformations,
\begin{equation}
Q = X W^Q, \quad K = X W^K, \quad V = X W^V,
\end{equation}
where $W^Q$, $W^K$, and $W^V$ are trainable matrices. The attention scores are then computed using a scaled dot-product,
\begin{equation}
A = \frac{\exp(QK^T / \sqrt{d})}{\sum \exp(QK^T / \sqrt{d})},
\end{equation}
where the summation runs over the index of $K$ and with $d$ being the dimensionality of the feature space. This mechanism  quantifies the contextual relationships between final-state objects by assigning dynamic importance weights. By evaluating each object’s significance, the model adaptively prioritises the most relevant features for classification. The self-attention mechanism further enhances performance by emphasising the objects that contribute most to the final prediction. 

The attention-weighted output for each particle token is given by
\begin{equation}
\mathcal{Z} = A \cdot V,
\end{equation}
preserving permutation invariance by involving a weighted sum across all tokens. Multiple attention heads are applied in parallel, enabling the model to extract diverse relational features in an efficient manner. Their outputs are subsequently concatenated and linearly transformed,
\begin{equation}
\widetilde{X} = \text{concatenate}(\mathcal{Z}_1, \dots, \mathcal{Z}_k) \times W,
\end{equation}
where $k$ is the number of heads and $W$ is a learnable projection matrix. A residual connection adds the original input to the transformed output, enabling the model to refine representations while maintaining stability. Since the output of each self-attention layer has the same dimensionality as the original input, as shown in equation \ref{eq:4.14}, multiple  layers can be stacked sequentially.

\begin{figure}
    \centering
    \includegraphics[width=0.5\linewidth]{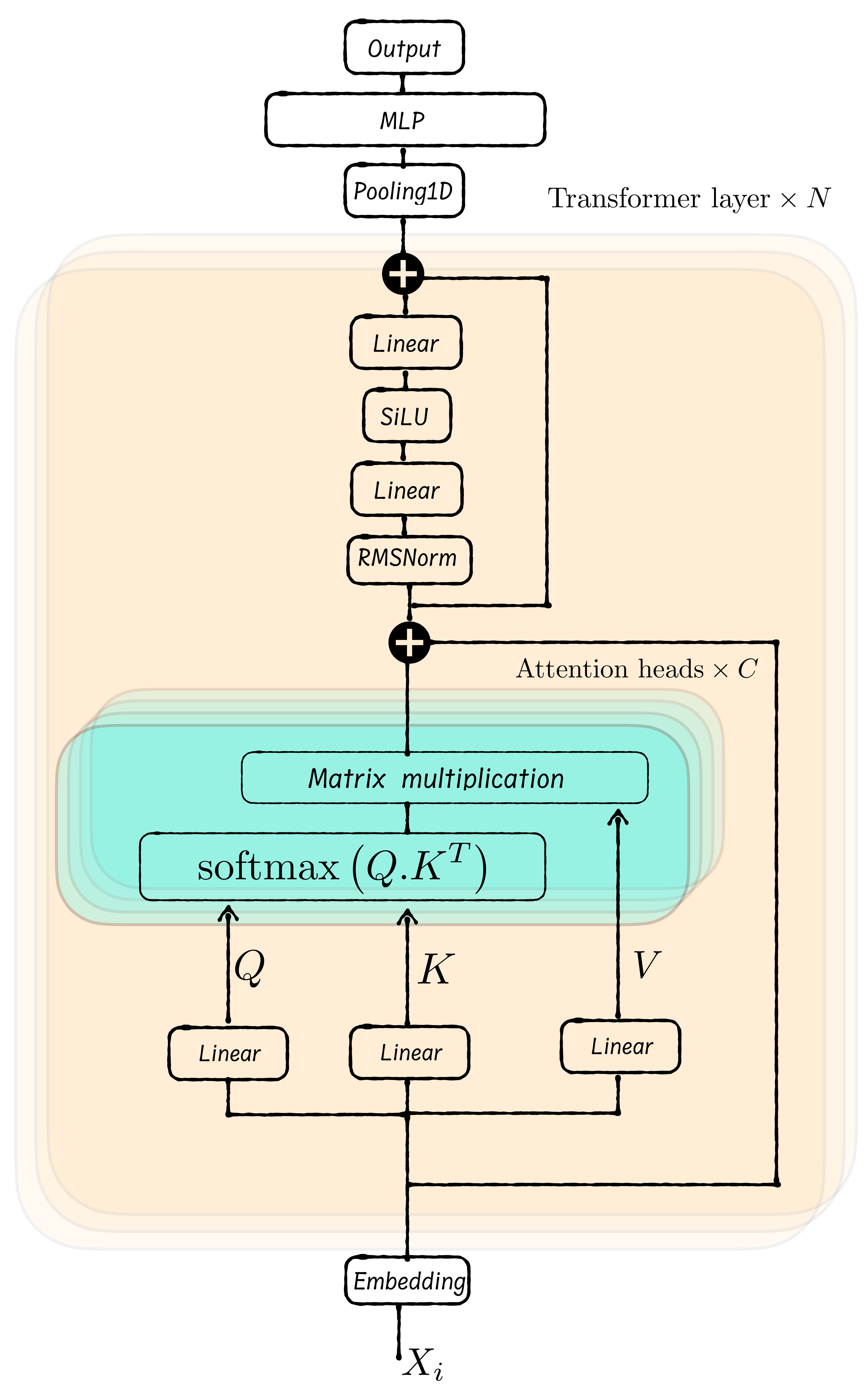}
    \caption{Schematic illustration of the used transformer network architecture. \label{fig:Transformer_network}}
\end{figure}

The transformer architecture employed in this work is illustrated in figure \ref {fig:Transformer_network} and begins with a series of fully connected embedding layers consisting of 512, 256 and 128 neurons. The resulting feature representation, of dimension $(9, 128)$, is passed through a normalisation layer and subsequently through a stack of 10 self-attention blocks. Each block contains 16 attention heads and incorporates a residual connection followed by Root Mean Square Layer Normalisation (RMSNorm) \cite{zhang2019root}. The output is then processed by two fully connected layers with Sigmoid-Weighted Linear Unit (SiLU) activations \cite{elfwing2018sigmoid}. After the final self-attention block, average pooling is applied across the particle dimension. The pooled vector is fed into a classification head composed of two dense layers with 128 and 64 units, respectively. Training is conducted over 21 epochs using a batch size of 128 and binary cross entropy as the loss function. Optimisation is performed using the AdamW algorithm with an initial learning rate of $0.0005$, adjusted during training via a cosine annealing scheduler~\cite{loshchilov2016sgdr}.

\begin{figure}
    \centering
    \includegraphics[width=0.9\linewidth]{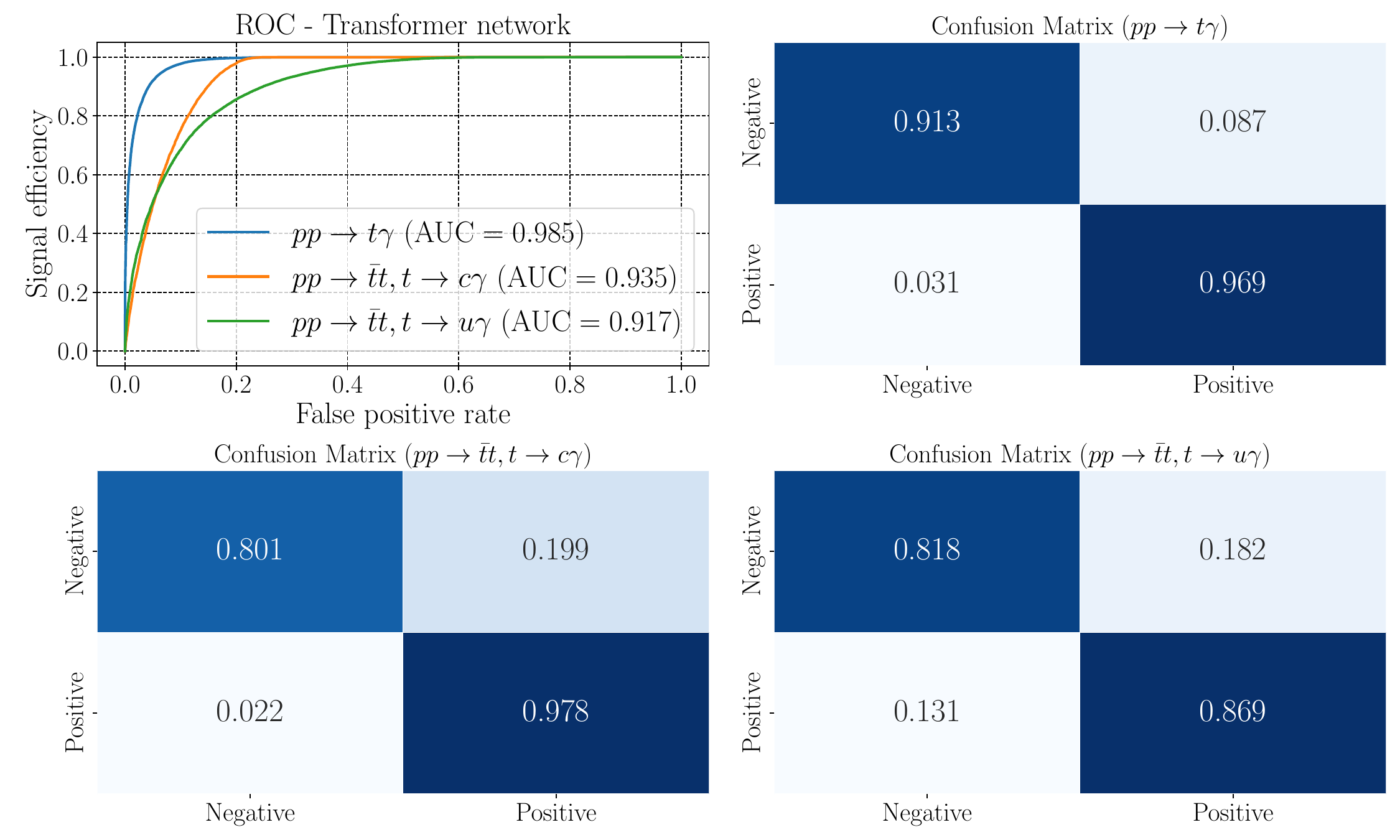}
    \caption{Classification performance of the transformer network used in our analysis, given as ROC curves (top-left) for $pp\to t\gamma$ (blue), $pp\to\bar{t}t$ with a $t\to u\gamma$ decay (orange) and $pp\to\bar{t}t$ with a $t\to c\gamma$ decay (green). The corresponding confusion matrices are given in the remaining panels.\label{fig:Transformer_results}}
\end{figure}

Figure~\ref{fig:Transformer_results} presents the classification performance across the three signal processes. The transformer achieves AUC values of 0.985 for FCNC single top production, 0.935 for top-pair production with a rare $t\to c\gamma$ decay and 0.917 for $t\to u\gamma$ and the confusion matrices highlight a similar behaviour. These results are consistent with, and in some cases outperform, those obtained with MLP and GAT models, demonstrating the transformer's strong ability to separate signal from background across complex event topologies.

%%%%%%%%%%%%%%%%%%%%%%%%%%%%%%
\section{Results}
\label{S:results:tqa}
%%%%%%%%%%%%%%%%%%%%%%%%%%%%%%
For each event, our different trained deep learning architectures yield a final output in the form of a non-negative real number in the range $[0, 1]$ representing the predicted probabilities of the event being signal-like. We remind that the classification performance of each model was previously evaluated using the ROC curves and confusion matrices presented in section~\ref{sec:DL}, and it was shown that the GAT and transformer architectures outperform the MLP model in terms of discriminative power. In our analysis, we define a signal region for each of the three channels (single top production, top pair production with one top decaying via $t \to u\gamma$ and top pair production with one top decaying via $t \to c\gamma$) by identifying the threshold value of $\mathcal{P}_{\rm sig}$ that maximises the signal significance defined in Eq.~\eqref{eq:significance}. In this procedure, we use the number of signal and background events obtained from true positive and true negative rates rescaled by the event weights from table~\ref{tab:xs}, and after assuming a baseline integrated luminosity of $\mathcal{L} = 100~{\rm fb}^{-1}$.

An important feature of the signal rates is that they scale proportionally to $|f_{tq}^\gamma|^2 + |h_{tq}^\gamma|^2$, as shown in sections~\ref{S:model} and ~\ref{sec:modelling}. This implies that our exclusion limits can be interpreted directly as bounds on this quadratic combination, and that we can simplify the parameter space by assuming that
\begin{equation}
    f_{tu}^\gamma = h_{tu}^\gamma\qquad\text{and}\qquad f_{tc}^\gamma = h_{tc}^\gamma.
    \label{eq:params:identities}
\end{equation}
The statistical interpretation of the results is finally performed using the $\text{CL}_s$ method, where the test statistic is based on the ratio of $p$-values
\begin{equation}
    \text{CL}_s = \frac{p_{\rm b+s}}{p_{\rm b}},
\end{equation}
with $p_{\rm b+s}$ and $p_{\rm b}$ denoting the $p$-values under the signal-plus-background and background-only hypotheses, respectively. A parameter point is hence considered excluded at $95\%$ confidence level if $\text{CL}_s < 0.05$. 

For each of the three network architectures and each of the three FCNC signal processes on which they have been trained (yielding a total of nine trained models), the number of signal events $N_s$ passing the final selection is defined as the sum of the contributions from single production (SP) and pair production (PP). In the training procedures, background events are always combined according to their weight (shown in Table \ref{tab:xs}) and the considered events account for the preselection cuts. Considering one flavour scenario at a time, we thus define
\begin{equation}
\begin{split}
    N_s(tu^\gamma) &\equiv N_s^{\rm SP}(f_{tc}^\gamma = 0) + N_s^{\rm PP}(f_{tc}^\gamma = 0), \\[0.2cm]
    N_s(tc^\gamma) &\equiv N_s^{\rm SP}(f_{tu}^\gamma = 0) + N_s^{\rm PP}(f_{tu}^\gamma = 0).
\end{split}
\end{equation}
Additionally, the expected number of background events is determined independently for each of the nine models, based on the corresponding trained network and the optimised signal selection. Exclusion limits are then derived using the \textsf{Pyhf} library~\cite{Heinrich:2021gyp}, employing the $\tilde{q}_\mu$ test statistic~\cite{Cowan:2010js} with a luminosity-dependent relative uncertainty for the background yield. The latter is fixed to be 10\% at $\mathcal{L} = 100~\text{fb}^{-1}$ and then scaled with luminosity according to Poisson statistics,
\begin{equation}
    \Delta N_b(\mathcal{L}) = 0.10 \times N_b(100~\text{fb}^{-1}) \times \sqrt{\frac{\mathcal{L}}{100~\text{fb}^{-1}}}.
\end{equation}
For each network architecture, the final ${\rm CL}_s$ value is taken to be the highest among the different signal-specific models when applied to the full signal combining the single and pair production channels, thus conservatively ignoring the potential gain that could stem from the combination of our three signal regions, which is hard to achieve without a proper assessment of the uncertainty correlations between the signal regions.

\begin{figure}[t]
    \centering
    \includegraphics[width=0.9\linewidth]{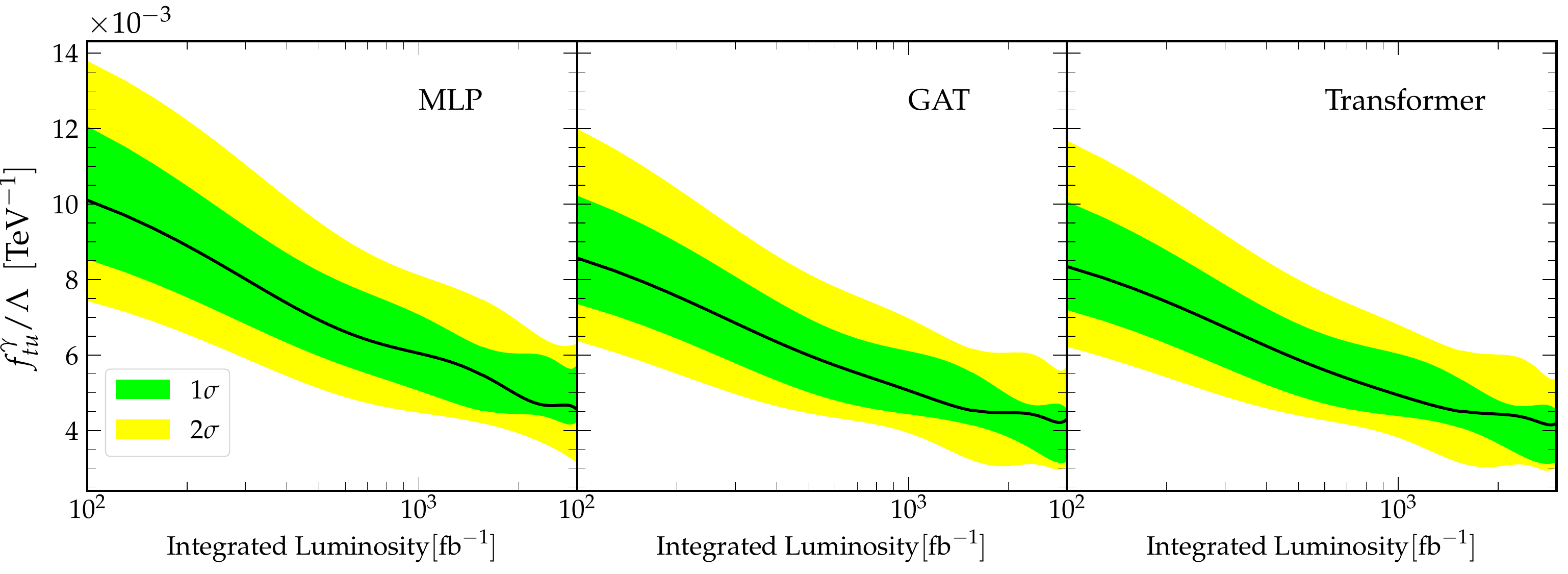}\vspace{.1cm}
    \includegraphics[width=0.9\linewidth]{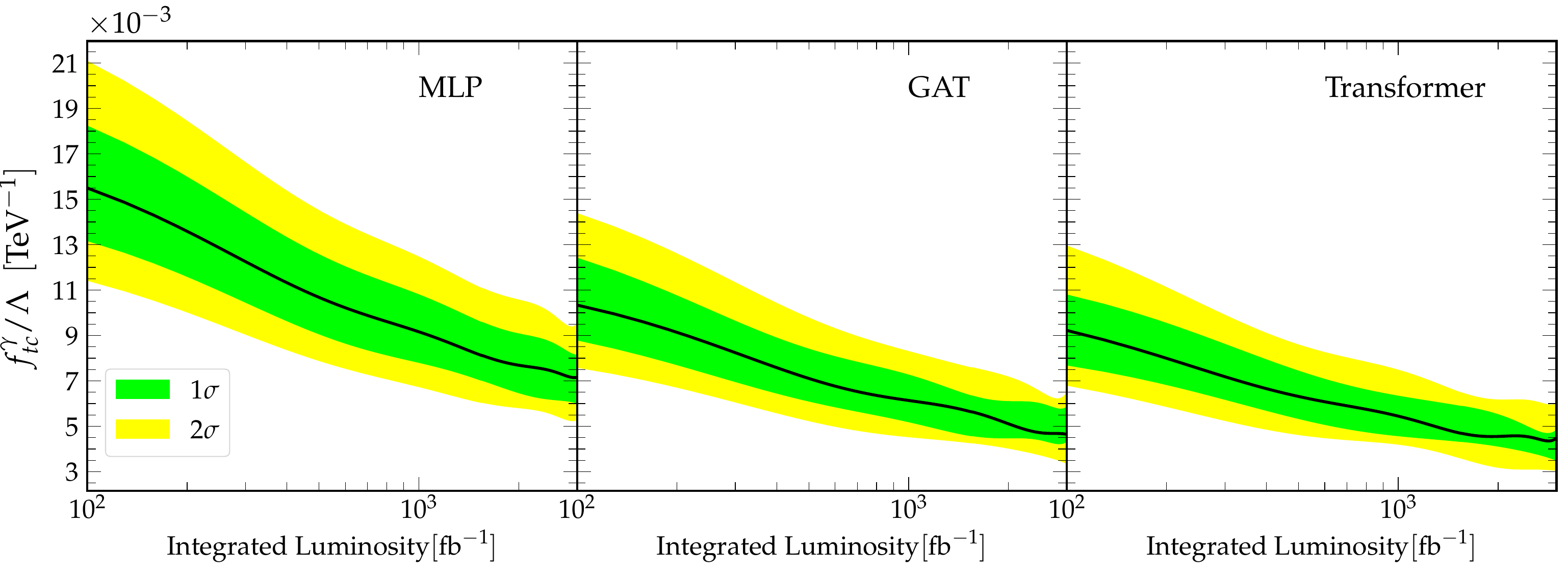}
    \caption{Expected 95\% CL exclusion limits at $\sqrt{s} = 13.6$~TeV on the FCNC couplings $f_{tu}^\gamma/\Lambda$ (top) and $f_{tc}^\gamma/\Lambda$ (bottom), shown as a function of the integrated luminosity. Results are given for  MLP (left), GAT (centre) and transformer (right) architectures, and the coloured bands indicate 1$\sigma$ and 2$\sigma$ uncertainties.\label{fig:exclusion:couplings}}
\end{figure}

Figure~\ref{fig:exclusion:couplings} presents the expected 95\% confidence level (CL) exclusions on the anomalous couplings $f_{tu}^\gamma/\Lambda$ (top row) and $f_{tc}^\gamma/\Lambda$ (bottom row) as a function of the integrated luminosity for each of the three considered networks, namely our MLP (left), GAT (centre) and transformer (right) architectures.  As anticipated, the bounds on FCNC couplings involving the up quark are stronger than those involving the charm quark. This trend is visible across all models, with the MLP yielding an improvement by a factor of approximately 1.6 for $f_{tu}^\gamma$ over $f_{tc}^\gamma$ while the GAT and transformer models show a milder factor of about $1.2-1.3$. This discrepancy is primarily driven by the kinematic and partonic structure of the LHC: the $u$-quark parton distribution function is significantly larger than that of the $c$-quark, leading to higher rates for the single top production mechanism mediated by $f_{tu}^\gamma$. Under the assumption $f_{tu}^\gamma = f_{tc}^\gamma$, the corresponding cross sections indeed differ by nearly a factor of 7 in favour of the $u$-quark channel, as shown in figure~\ref{fig:xsec:SP}. However, this difference is partially offset by the better sensitivity to charm-quark FCNC in the pair production channel where the top decay $t \to c\gamma$ leads to a more distinctive final state than the $t \to u\gamma$ decay. This balance between production rate and classifier efficiency helps explain why the transformer and GAT models display a reduced gap between the two flavour scenarios compared to the MLP.

\begin{table}
\setlength\tabcolsep{7pt}\renewcommand{\arraystretch}{1.2}
\begin{center}
\begin{tabular}{l ccc}
Constrained quantity & MLP & GAT & Transformer \\
\toprule
$f_{tu}^\gamma/\Lambda = h_{tu}^\gamma/\Lambda~~[\times 10^{-3}]$ & $9.2^{+1.6}_{-1.4}~(6.6^{+1.2}_{-0.8})$ & $7.6^{+1.4}_{-1.3}~(5.7^{+0.6}_{-1.0})$ & $7.3^{+1.2}_{-1.1}~(5.4^{+0.8}_{-0.8})$ \\ [0.4em]
$|C_{uB}^{31} + C_{uW}^{31}|/\Lambda^2~~[\times 10^{-8}]$ & $15.0^{+2.7}_{-2.3}~(10.9^{+1.9}_{-1.3})$ & $12.4^{+2.3}_{-2.1}~(9.3^{+1.0}_{-1.7})$ & $11.9^{+2.0}_{-1.9}~(8.8^{+1.3}_{-1.4})$ \\ [0.4em]
${\rm BR}(t\to u\gamma)~~[\times 10^{-6}]$ & $10.2^{+4.0}_{-2.9}~(5.4^{+2.1}_{-1.2})$ & $7.1^{+2.8}_{-2.2}~(4.0^{+0.9}_{-1.4})$ & $6.6^{+2.4}_{-1.9}~(3.6^{+1.1}_{-1.0})$ \\[.2cm] 
\toprule
$f_{tc}^\gamma/\Lambda = h_{tc}^\gamma/\Lambda~~[\times 10^{-3}]$ & $14.2^{+2.7}_{-2.0}~(10.6^{+1.8}_{-1.6})$ & $9.7^{+1.9}_{-1.5}~(7.2^{+1.2}_{-1.1})$ & $8.7^{+1.7}_{-1.3}~(6.2^{+1.3}_{-0.8})$ \\ [0.4em]
$|C_{uB}^{32} + C_{uW}^{32}|/\Lambda^2~~[\times 10^{-8}]$ & $23.4^{+4.3}_{-3.3}~(17.4^{+2.9}_{-2.6})$ & $15.6^{+3.1}_{-2.4}~(11.8^{+1.9}_{-1.9})$ & $14.2^{+2.7}_{-2.1}~(10.2^{+2.1}_{-1.3})$ \\ [0.4em]
${\rm BR}(t\to c\gamma)~~[\times 10^{-6}]$ & $24.7^{+10.6}_{-6.5}~(13.7^{+5.0}_{-3.8})$ & $11.4^{+4.9}_{-3.2}~(6.4^{+2.1}_{-1.9})$ & $9.2^{+3.8}_{-2.5}~(4.7^{+2.2}_{-1.1})$ \\[.2cm]
\end{tabular}
\end{center}
  \caption{Expected 95\% CL exclusions (with associated $1\sigma$ uncertainties) on top-photon FCNC, for integrated luminosities of 139~fb$^{-1}$ and the LHC Run~3 target of 500~fb$^{-1}$ (in parentheses). Results are presented for each of the three network architectures considered and in terms of anomalous FCNC couplings (first row in each panel), relevant combination of SMEFT Wilson coefficients (second row in each panel) and top rare branching ratios (last row in each panel). \label{tab:exclusions}}  
\end{table}

In fact, the transformer and GAT models significantly outperform the MLP in both the up and charm FCNC channels. At an integrated luminosity of 139~fb$^{-1}$, the GAT and transformer architectures exclude $f_{tc}^\gamma/\Lambda$ values down to $(8.7-14.2) \times 10^{-3}~\text{TeV}^{-1}$ and $f_{tu}^\gamma/\Lambda$ values down to $(7.3-9.2) \times 10^{-3}~\text{TeV}^{-1}$, the exact value depending on the deep learning model (see table~\ref{tab:exclusions}). These improvements amount to a relative gain in sensitivity of about $30\%-70\%$ over the MLP thanks to the superior ability of attention-based architectures to capture non-trivial correlations between final-state particles. Furthermore, these exclusions scale favourably with luminosity. At the HL-LHC target of 3~ab$^{-1}$, all models predict exclusion of couplings as small as approximately $2\times 10^{-3}~\text{TeV}^{-1}$, showing that such analyses could reach unprecedented precision. Predictions for a luminosity of 500~fb$^{-1}$ that is expected for the LHC Run~3 are also presented in table~\ref{tab:exclusions}. Our results hence reinforce the value of incorporating graph-based and attention-based architectures in collider searches involving subtle signals with complex final states.

\begin{figure}[t]
    \centering
    \includegraphics[width=0.9\linewidth]{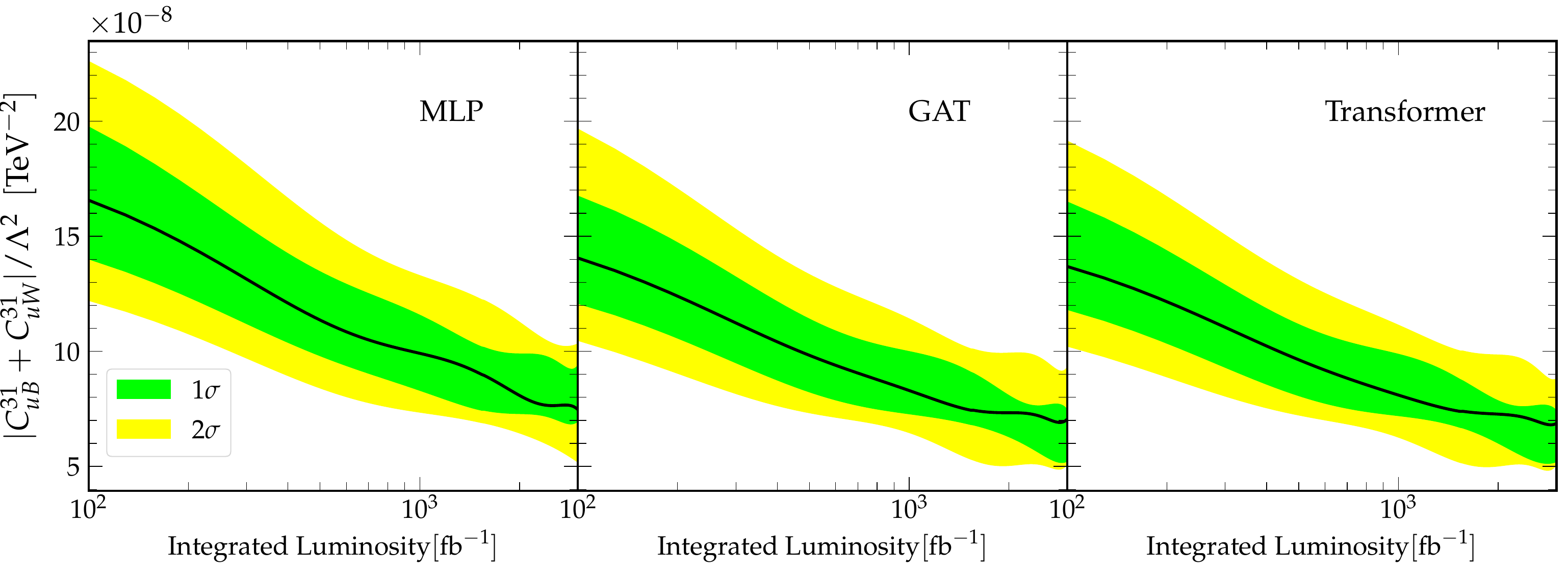}\\
    \includegraphics[width=0.9\linewidth]{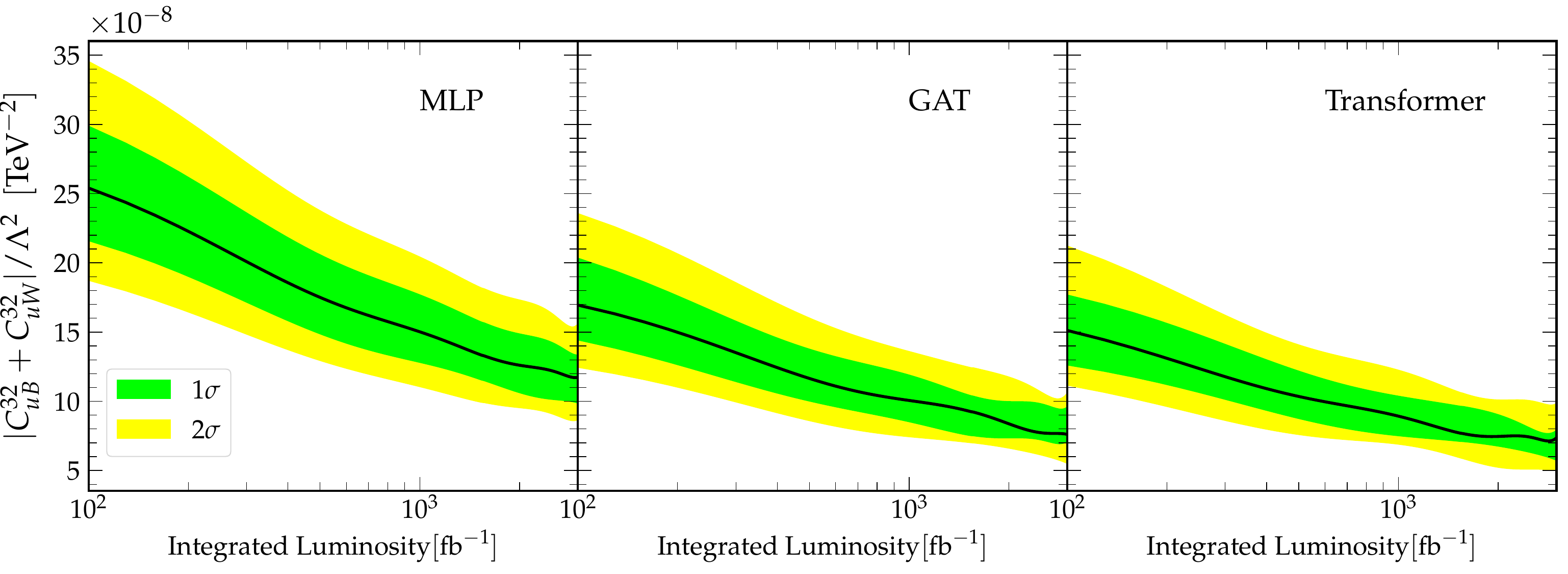}
    \caption{Same as figure \ref{fig:exclusion:couplings} but for the relevant combination of SMEFT Wilson coefficients.\label{fig:exclusions:WCs}}\vspace{.1cm}
    \includegraphics[width=0.9\linewidth]{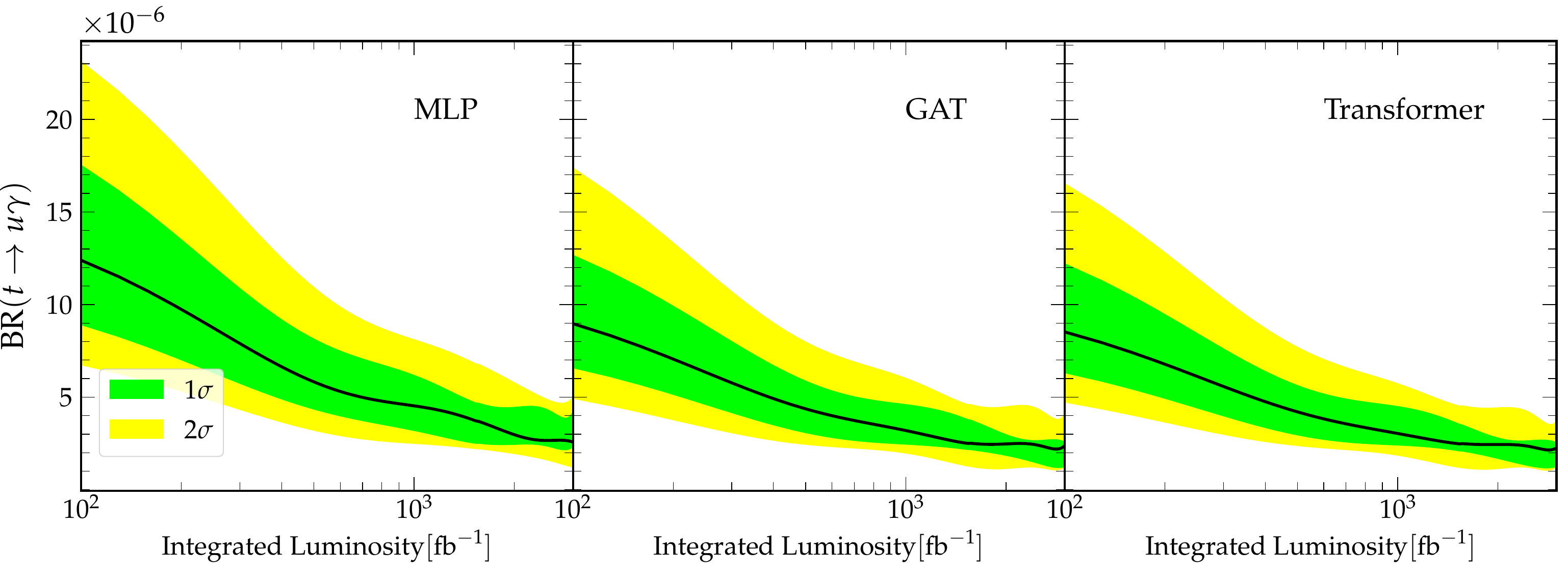}\\
    \includegraphics[width=0.9\linewidth]{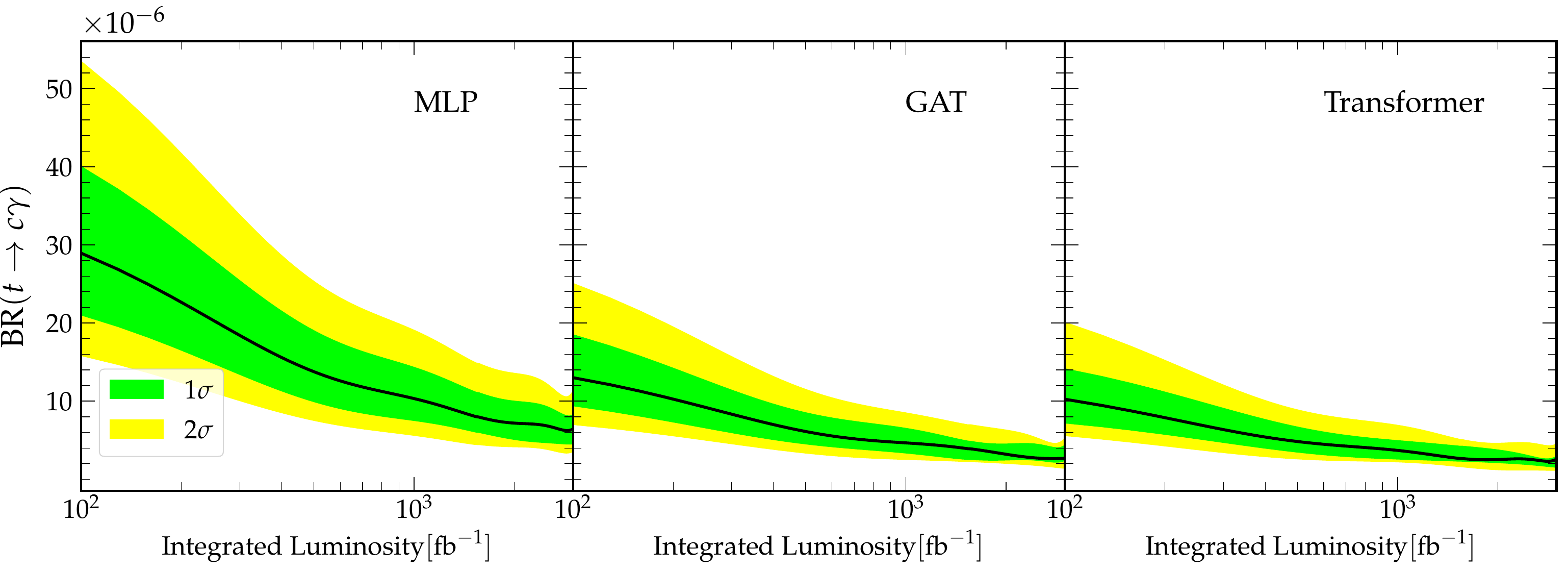}
    \caption{Same as figure \ref{fig:exclusion:couplings} but for the associated rare top branching ratios.\label{fig:exclusions:BRs}}
\end{figure}

We now turn to a discussion of the implications for the Wilson coefficients $C_{uB}$ and $C_{uW}$. Using Eq.~\eqref{eq:WCs:params} along with the simplifying identities defined in Eq.~\eqref{eq:params:identities}, we obtain the following relations,
\begin{equation}
    \frac{1}{\Lambda^2} \left| C_{uB}^{31} + C_{uW}^{31} \right| = 2\sqrt{2} \frac{f_{tu}^\gamma}{m_t \Lambda}, \qquad 
    \frac{1}{\Lambda^2} \left| C_{uB}^{32} + C_{uW}^{32} \right| = 2\sqrt{2} \frac{f_{tc}^\gamma}{m_t \Lambda},
\end{equation}
with similar expressions holding for the complex conjugates quantities $|(C_{uB}^{a3} + C_{uW}^{a3})^*|$. The corresponding exclusions on these Wilson coefficients are presented in figure~\ref{fig:exclusions:WCs}. At an integrated luminosity of 139~fb$^{-1}$, values of $|C_{uB}^{3a} + C_{uW}^{3a}|/\Lambda^2$ in the range of $11.4 - 23.4 \times 10^{-8}~{\rm TeV}^{-1}$ are excluded at the 95\% confidence level, depending on the network architecture and the flavour index $a$. As observed in previous results, the strongest bounds involving couplings to the first-generation quark ($a=1$), obtained with attention-based models, are approximately 1.2 to 1.6 times more stringent than those involving second-generation couplings ($a=2$).

We conclude this section by representing the bounds as limits on the FCNC branching ratios of the top quark, figure~\ref{fig:exclusions:BRs} showing the expected exclusions for ${\rm BR}(t\to u\gamma)$ (top row) and ${\rm BR}(t\to c\gamma)$ (bottom row). At $\mathcal{L} = 139~{\rm fb}^{-1}$, the MLP architecture yields expected limits of ${\cal O}(10^{-5})$, with ${\rm BR}(t \to u\gamma) \lesssim 1.0 \times 10^{-5}$ and ${\rm BR}(t \to c\gamma) \lesssim 2.4 \times 10^{-5}$. As already found previously, these limits improve when using the GAT or transformer models, this time by factors of 1.5 to 2.6. Subsequently, at the HL-LHC branching ratios as low as ${\cal O}(10^{-6})$ are potentially excludable at 95\% CL, particularly when using our transformer-based classifier.

%%%%%%%%%%%%%%%
\section{Summary and conclusions}
\label{S:conclusions}
%%%%%%%%%%%%%%%
The LHC, with its high top quark production rate, provides a unique environment for probing rare top quark decays, thus allowing stringent constraints to be placed on models of new physics and offering significant discovery potential in future data-taking campaigns. In this study, we have explored the use of deep learning techniques to enhance the sensitivity to FCNC interactions involving the top quark and a photon. We focused on two complementary processes: the rare decay $t \to q\gamma$ in top pair production and the anomalous single top production mechanism $qg \to t\gamma$. These channels are specifically sensitive to anomalous top FCNC couplings induced by linear combinations of several dimension-six operators in the SMEFT framework which impact both the total production rates and the kinematic structure of the final states.

We began by evaluating the sensitivity of a traditional cut-based analysis % using a set of manually optimised kinematic selections. Although this strategy 
to obtain baseline results, the performance being obviously limited by the analysis inability to capture complex correlations among observables. % The resulting signal significances were modest, underscoring the need for more advanced classification approaches.
To address this, we next employed three deep learning architectures to perform a supervised binary classification task designed to separate signal from background. Among these, MLPs exhibited the lowest performance, with an AUC of 0.928 for the single top channel and approximately 0.78 for top pair production, this resulting from solely relying on event-level kinematic variables. An analysis of feature importance showed that the classification performance was dominated by a small subset of inputs including the $H_T$, $\Delta R(b,\ell)$ and $p_T^\gamma$ observables, which together account for nearly 80\% of the network’s predictive power.

Transformer networks demonstrated the highest classification performance, achieving AUC values of 0.985 and about  0.93 for single top and top pair production respectively. Their strength lies in their ability to process particle-level input as unordered sets (or clouds), preserving the relational and spatial structure of the event. Each particle is treated as a token, and the self-attention mechanism dynamically identifies the most relevant features across the event, thus allowing the network to construct rich and hierarchical representations of the data that incorporate both local and global information. The GAT model achieved instead intermediate performance, with AUC values of 0.973 for the single top channel and approximately 0.92 for top pair production. GATs benefit from a graph-based representation of the final state, in which particles are treated as nodes, and the attention mechanism focuses on the most relevant neighbours during message passing. This subsequently effectively captures the event topology and enhances the model's ability to discriminate between signal and background.

The improved classification capabilities of the GAT and transformer networks translated directly into more stringent bounds on the FCNC couplings. Assuming an integrated luminosity of $\mathcal{L} = 100~{\rm fb}^{-1}$ and a $10\%$ background uncertainty, we found that the transformer and GAT models yield expected limits on the anomalous couplings $f_{tu}^\gamma/\Lambda$ and $f_{tc}^\gamma/\Lambda$ that are stronger by a factor of approximately $1.3-1.7$ compared to the MLP. %, that itself outperforms the performance of our simpler cut-based analysis.
At $\mathcal{L} = 139~{\rm fb}^{-1}$, values of $f_{tu}^\gamma/\Lambda$ down to $7.3-9.2 \times 10^{-3}~{\rm TeV}^{-1}$ and $f_{tc}^\gamma/\Lambda$ down to $8.7-14.2 \times 10^{-3}~{\rm TeV}^{-1}$ can hence be excluded, which will improve further at the HL-LHC where coupling values as small as $2 \times 10^{-3}~{\rm TeV}^{-1}$ may be probed. The exclusions on the corresponding FCNC branching ratios follow a similar trend. At $\mathcal{L} = 139~{\rm fb}^{-1}$, we find that the MLP excludes ${\rm BR}(t\to u\gamma)$ and ${\rm BR}(t\to c\gamma)$ down to about $1.0 \times 10^{-5}$ and $2.4 \times 10^{-5}$, respectively, with the GAT and transformer architectures improving these bounds by factors of $1.5-2.6$. At the HL-LHC, branching fractions as small as $10^{-6}$ could eventually be excluded at the 95\% confidence level using transformer-based classification.

In summary, our results demonstrate that % the power of incorporating deep learning approaches into collider analyses, particularly for rare top quark decay processes. With 
with the vast increase in statistics expected from future LHC runs, traditional machine-learning methods such as MLPs will become increasingly suboptimal. Our findings suggest that attention-based architectures, such as transformers and GATs, offer significant potential for improving the sensitivity to new physics signatures in top quark processes at the LHC and beyond.

%%%%%%%%%%%%%%%%%%%%%%%%%%%%%%%%
\section*{Acknowledgments}%%%%%%%%
%%%%%%%%%%%%%%%%%%%%%%%%%%%%%%%%
The work of SKG has been supported by SERB, DST, India through grant TAR/2023/000116. SKG acknowledges the Manipal Centre for Natural Sciences, Centre of Excellence, Manipal Academy of Higher Education (MAHE) for facilities and support. AH is funded by the Grant Number 22H05113 from the ``Foundation of Machine Learning Physics'', the ``Grant in Aid for Transformative Research Areas''  22K03626 and the Grant-in-Aid for Scientific Research (C). The work of AJ is supported by the Institute for Basic Science (IBS) under Project Code IBS-R018-D1. BF has been supported by Grant ANR-21-CE31-0013 from the \emph{Agence Nationale de la Recherche} (France).

%%%%%%%%%%%%%%%%%%%%%%%%%%%%
\bibliographystyle{JHEP}
\bibliography{main}

%%%%%%%%%%%%%%%%%%%%%%%%%%%%%%%%
\end{document}